\documentstyle[12pt,a4]{article}

\input{amssym.def}
\input{amssym.tex}

\newfont{\Bbbtw}{msbm10 scaled 1200}
\newfont{\fraktw}{eufm10 scaled 1200}
\newfont{\frakte}{eufm10}
\newfont{\ninBbb}{msbm9} 
\newfont{\eigBbb}{msbm8} 
\newfont{\sevBbb}{msbm7} 
\newfont{\ninfrak}{eufm9} 
\newfont{\eigfrak}{eufm8} 
\newfont{\sevfrak}{eufm7}

\newcommand{\smsupset}{\,\raisebox{0.06em}{${\scriptstyle \supset}$}\,}

\newcommand{\smcap}{\,\raisebox{0.06em}{${\scriptstyle \cap}$}\,}
\newcommand{\bm}[1]{\mbox{\boldmath{$#1$}}}

\newcommand{\cg}{\mbox{\fraktw g}}

\begin{document}
\makebox[13.5cm][r]

\vspace{1.5cm}

\begin{center}
{\bf Lie Superalgebras and the Multiplet Structure \\
     of the Genetic Code II: Branching Schemes}

\vspace{1.5cm}
\renewcommand{\thefootnote}{\fnsymbol{footnote}}

 {\large Michael Forger$\,^1$ and Sebastian Sachse$\,^2$
         \footnote{Work supported by FAPESP (Funda\c{c}\~ao de Amparo
                   \`a Pesquisa do Estado de S\~ao Paulo) and CNPq
                   (Conselho Nacional de Desenvolvimento Cient\'{\i}fico
                   e Tecnol\'ogico), Brazil}}
 \\[5mm]
      $^1\,$ {\em Departamento de Matem\'atica Aplicada,} \\
      $^2\,$ {\em Departamento de Matem\'atica,} \\[1mm]
 {\em Instituto de Matem\'atica e Estat\'{\i}stica, \\
      Universidade de S\~ao Paulo, \\
      Cx.\ Postal 66281, BR--05315-970 S\~ao Paulo, S.P., Brazil} \\[1cm]
\end{center}
\renewcommand{\thefootnote}{\arabic{footnote}}
\setcounter{footnote}{0}

\thispagestyle{empty}

\vspace{1cm}

\begin{abstract}
\noindent
Continuing our attempt to explain the degeneracy of the genetic code using
basic classical Lie superalgebras, we present the branching schemes for
the typical codon representations (typical $64$-dimensional irreducible
representations) of basic classical Lie superalgebras and find three
schemes that do reproduce the degeneracies of the standard code, based
on the ortho\-symplectic algebra $\mbox{\frakte osp}(5|2)$ and differing
only in details of the symmetry breaking pattern during the last step.
\end{abstract}

\begin{flushright}
 \parbox{12em}
 { \begin{center}
 Universidade de S\~ao Paulo \\
 RT-MAP 99/01 \\
 April 1999
 \end{center} }
\end{flushright}

\newpage

\section{Introduction}

In the context of the project proposed by Hornos \& Hornos \cite{HH} which
aims at explaining the degeneracy of the genetic code as the result of a
symmetry breaking process, we have carried out a systematic analysis of
the possibility to implement this idea by starting out from a typical
codon representation (typical $64$-dimensional irreducible representation)
of a basic classical Lie superalgebra, rather than a codon representation
($64$-dimensional irreducible representation) of an ordinary simple Lie
algebra. The investigation of such an algebraic approach to the genetic
code using alternative concepts of symmetry such as supersymmetry, where
ordinary Lie algebras are replaced by Lie superalgebras, has already been
suggested in the original paper \cite{HH} -- except for the restriction
to basic classical Lie superalgebras (a particular class of simple Lie
superalgebras) and to typical representations (a particular class of
irreducible representations): only under this restriction, which is
of a technical nature, does there exist a sufficiently well developed
mathematical theory, due to Kac \cite{K1,K2}, to allow for the kind of
analysis that is necessary to carry out such a program. As a first step,
we have in a previous paper \cite{FS} presented a complete classification
of all typical codon representations of basic classical Lie superalgebras:
there are altogether $18$ such representations involving $12$ different Lie
superalgebras. Our goal in the present paper is to analyze all possible
branching schemes that can be obtained from these representations with
regard to their capability of reproducing the degeneracy of the genetic
code, following the strategy used in ref.\ \cite{HH} and explained in
detail in ref.\ \cite{HHF}, but with one essential restriction:
supersymmetry will be broken right away, in the very first step.

To motivate this assumption, note that the distribution of multiplets
found in the genetic code today does not appear to correspond to the
kind of scheme one would expect from the representation theory of Lie
superalgebras. Thus if some kind of supersymmetry has been present at
the very beginning of the evolution of the genetic code, it must have
been broken. Moreover, it does not seem plausible to us that this
breaking should have occurred only in the last step of the process,
where the phenomenon of ``freezing'' would have been able to prevent
a complete breakdown (see ref.\ \cite{HHF} for more details). But if
supersymmetry has been broken before, then there is mathematically
no loss of generality in assuming that it has been broken in the very
first step, because as soon as we may exclude freezing, symmetry
breaking through chains of subalgebras that differ only in the order
in which the successive steps are performed (such as $\; \cg \smsupset
\cg_1 \smsupset \cg_1 \smcap \cg_2$ \linebreak and $\; \cg \smsupset \cg_2
\smsupset \cg_1 \smcap \cg_2$) will lead to the same end result.

\section{The first step: Breaking the supersymmetry}

With the above picture in mind and using the fact that among the semisimple
ordinary Lie algebras which are subalgebras of a given basic classical Lie
super\-algebra $\cg$, there is a unique maximal one, namely the semisimple
part $\cg_{\bar{0}}^{\rm ss}$ of its even part $\cg_{\bar{0}}$, our task
for the first step of the symmetry breaking process is to compute, for
each of the $18$ codon representations of the $12$ basic classical Lie
super\-algebras found in ref.\ \cite{FS}, its branching into irreducible
representations \linebreak under restriction from $\cg$ to $\cg_{\bar{0}}^%
{\rm ss}$. There are two different methods for doing this. One consists
in computing all weight vectors that result from the action of products
of generators associated with the negative odd roots on the highest
weight vector, where every negative odd root appears at most once in
such a product: these are the candidates for highest weight vectors of
irreducible representations of $\cg_{\bar{0}}^{\rm ss}$ that appear in
the direct decomposition of the original codon representation of $\cg$.
The problem is to decide which of these representations really appear,
and with what multiplicity. Although there is an explicit formula for
calculating such multiplicities, due to Kac and Kostant, the procedure
involves a summation over the Weyl group and is cumbersome to apply in
practice. Therefore, we shall, following common usage, adopt the other
method, which is based on the use of Young superdiagrams -- a generalization
of the usual Young diagrams from ordinary Lie algebras to Lie superalgebras.

In order to understand how this technique works, it is useful to recall
how Young diagrams arise in the representation theory of ordinary simple Lie
algebras. \linebreak Given a simple Lie algebra $\cg_0$, consider the first
fundamental representation of $\cg_0$, i.e., the irreducible representation
of $\cg_0$ with highest weight equal to the first fundamental weight, denoted
in what follows by $D$. Alternatively, we may characterize $D$ as the lowest-%
dimensional (non-trivial) irreducible representation of $\cg_0$: for
the matrix Lie algebras $\mbox{\fraktw sl}(n)$, $\mbox{\fraktw so}(n)$
and $\mbox{\fraktw sp}(n)$, it is simply the $n$-dimensional defining
representation. The basic idea is now to look at all \mbox{tensor} powers
$D^{\otimes p}$ of $D$ and reduce them into their irreducible constituents.
This \linebreak reduction is achieved by considering symmetric tensors,
antisymmetric tensors and, more generally, tensors of mixed symmetry type.
In fact, permutation of the factors induces a representation of the
symmetric group $S_p$ on the representation space of $D^{\otimes p}$
and this action of $S_p$ commutes with that of $\cg_0$, so that both
actions can be simultaneously decomposed into irreducible constituents.
More precisely, this is achieved by combining them into a ``joint action''%
\footnote{The concept of ``joint action'' used here can be formulated in
mathematically rigorous terms by introducing the connected, simply connected,
simple Lie group $G_0$ corresponding to $\mbox{\frakte g}_0$ and considering
$D$ and $D^{\otimes p}$ as representations of $G_0$; then the joint action of
$S_p$ and $\mbox{\frakte g}_0$ corresponds to a representation of the direct
product $\, S_p \times G_0$.} and then performing a decomposition into
irreducible constituents in the usual sense: each of these has the
property that its multiplicity as a representation of $S_p$ equals
its dimension as a representation of $\cg_0$ and its multiplicity as
a representation of $\cg_0$ equals its dimension as a representation
of $S_p$. The usefulness of this approach stems from an important theorem
of Weyl which states that any irreducible representation of the classical
Lie algebras $\mbox{\fraktw sl}(n)$ and $\mbox{\fraktw sp}(n)$, as well as
any tensorial irreducible representation of the classical Lie algebras
$\mbox{\fraktw so}(n)$, can be obtained in this way.\footnote{An
irreducible representation of $\mbox{\frakte so}(n)$ of highest
weight $(l_1,\ldots,l_{r-1},l_r)$ is tensorial, or non-spinorial,
if $l_r$ is even for $\, n = 2r+1 \,$ odd ($B$-series) and if
$\, l_{r-1} + l_r \,$ is even for $\, n = 2r \,$ even ($D$-series).}
Therefore, a Young diagram of $p$ boxes, which originally stands
for an irreducible representation of the symmetric group $S_p$,
also determines an irreducible representation of $\cg_0$ contained
in $D^{\otimes p}$. In the case of $\mbox{\fraktw sl}(n)$, the latter
is simply obtained by considering tensors of a specific symmetry type,
given by the projection operator of symmetrizing along the rows and
antisymmetrizing along the columns of the corresponding standard Young
tableau \cite{Boe,Ham}, whereas in the case of $\mbox{\fraktw sp}(n)$
and $\mbox{\fraktw so}(n)$, the existence of invariant bilinear forms
for $D$ (antisymmetric for $\mbox{\fraktw sp}(n)$ and symmetric for
$\mbox{\fraktw so}(n)$) implies that this operation alone is not
sufficient to produce an irreducible representation: here, a given
Young diagram stands for tensors of the corresponding symmetry type which
in addition are totally traceless with respect to the pertinent bilinear
form, that is, traceless in all indices in which they are antisymmetric
in the case of $\mbox{\fraktw sp}(n)$ and traceless in all indices in
which they are symmetric in the case of $\mbox{\fraktw so}(n)$.

The rules for constructing Young tableaux and diagrams can be
extended in such a way as to also cover spinorial representations
of $\mbox{\fraktw so}(n)$. To this end,
one must include the spinor representation(s), i.e., the standard spinor
representation $S$ of highest weight $(0,\ldots,0,1)$ if $n$ is odd and
the two chiral spinor representations $S^+$ and $S^-$, of highest weight
$(0,\ldots,0,1,0)$ and $(0,\ldots,0,0,1)$, respectively, if $n$ is even:
this turns out to be sufficient because according to a modified form
of Weyl's theorem, an arbitrary irreducible representation of
$\mbox{\fraktw so}(n)$ can be obtained as a subrepresentation of the
representation $\, D^{\otimes p} \otimes S \,$ if $n$ is odd and of
one of the two representations $\, D^{\otimes p} \otimes S^+ \,$ or
$\, D^{\otimes p} \otimes S^- \,$ if $n$ is even, for adequate~$p$.
Therefore, it is convenient to introduce generalized Young tableaux
and diagrams containing ``spinor'' or ``half'' boxes, one at the
beginning of each row\footnote{The property of having only one
spinor box per row reflects the fact that the spinor representation(s)
appear only once in the tensor product, so that in particular, there is
no problem with symmetrization or antisymmetrization of spinor indices.}
and characterized by inserting the letter ``s'' into each of them, as
well as a possible ``negative'' last row instead of the usual ``positive''
one when $n$ is even, thus allowing to distinguish between the two
chiralities for the spinors. For a summary of the conventions that
we shall follow, the reader is referred to the Appendix of ref.\
\cite{MSS}.

An important point to be noticed is that although different (generalized)
Young diagrams correspond to different irreducible representations of the
permutation group $S_p$, they may very well describe the same irreducible
representation \linebreak of $\cg_0$: thus the characterization of
irreducible representations of $\cg_0$ by (generalized) Young diagrams
is ambiguous. In order to remove this ambiguity, one introduces {\em
modification rules} which allow to reduce every (generalized) Young
diagram to its {\em standard} form, as explained, for instance,
in \cite{CK}: this is done in such a way that every irreducible
representation corresponds to precisely one standard (generalized)
Young diagram.

The technique of (generalized) Young tableaux and Young diagrams for
characterizing irreducible representations has been extended from the
classical simple Lie algebras to the special linear and orthosymplectic
Lie superalgebras, giving rise to {\em Young supertableaux} and {\em Young
superdiagrams}, which we shall distinguish from their non-supersymmetric
counterparts by the insertion of a diagonal line across each box. They
describe typical representations as well as atypical ones. \linebreak
As in the non-supersymmetric case, several different Young superdiagrams
may provide the same irreducible representation, and modification rules
are needed to remove the ambiguity: they serve to reduce a Young super\-%
diagram to its \linebreak {\em standard} form. For an atypical representation,
this procedure is still not unambiguous, leading to different standard Young
super\-diagrams describing the same representation, whereas for a typical
representation, the corresponding Young super\-diagram can be constructed
directly from its highest weight, and conversely, the Kac-Dynkin labels of
the highest weight may be read off from the Young super\-diagram. Note that
fixing the highest weight includes fixing the Kac-Dynkin label $l_s$ of the
simple odd root, which for type I Lie superalgebras can take continuous
values: the corresponding irreducible representation will in that case
carry an additional continuous parameter. Its dimension and its branching
rules under reduction from $\cg_0$ to $\cg_{\bar{0}}^{\rm ss}$ will however
not depend on the value of $l_s$ which in \cite{FS} had remained unspecified,
except for the constraints imposed by requiring typicality of the
representation. Here, we shall make a choice for $l_s$ that leads to
the simplest possible Young superdiagram which is consistent with these
constraints; this value, together with the resulting Young superdiagram,
is specified in Tables 1-3.

For the special linear Lie superalgebras $\mbox{\fraktw sl}(m\,|\,n)$,
the procedure of constructing irreducible representations from Young
superdiagrams is straightforward. The main difference from the case
of the special linear Lie algebras $\mbox{\fraktw sl}(n)$ is that
the process of symmetrization and antisymmetrization involved in
the definition of the Young idempotents that project onto a tensor
of a specific symmetry type must now be understood in the appropriate
supersymmetric or graded sense: symmetrization or antisymmetrization of
two fermionic indices involves an extra minus sign to take into account
the anticommuting character of these variables. This implies that there
no longer exists an invariant totally antisymmetric tensor of top degree
(invariant volume or $\epsilon$-tensor), so the irreducible representations
$D$ and $\bar{D}$ become independent; therefore Young superdiagrams for
$\mbox{\fraktw sl}(m\,|\,n)$ are in general composed of ``undotted''
and ``dotted'' boxes, as happens in the case of Young diagrams for
$\mbox{\fraktw gl}(n)$. For the applications needed in this paper,
however, we shall find it sufficient to use the conventional type
of Young superdiagram containing only ``undotted'' boxes.

The relation between such Young superdiagrams and Kac-Dynkin labels
of irreducible representations for $\mbox{\fraktw sl}(m\,|\,n)$ can
be summarized as follows. First, recall that an ordinary Young diagram,
characterized by a nonincreasing sequence $\, b_1 \geq \ldots \geq b_r \,$
of positive integers giving the lengths of its $r$ rows\footnote{It is to
be understood that $\, b_r > 0 \,$ but $\, b_i = 0 \,$ if $\, i > r$.},
will be an allowed Young diagram for $\mbox{\fraktw sl}(n)$ if and only if
$\, r \leq n$; in this case it will describe an irreducible representation
of $\mbox{\fraktw sl}(n)$ with Dynkin labels $\, l_1,\ldots,l_{n-1} \,$
given by
\begin{equation} \label{eq:DLSL}
 l_i~=~b_i - b_{i+1} \qquad \mbox{for}~i = 1,\ldots,n-1~, \\[1mm]
\end{equation}
Similarly, according to ref.\ \cite{BMR}, a Young super\-diagram containing
only ``undotted'' boxes, characterized by nonincreasing sequences $\, b_1
\geq \ldots \geq b_r \,$ and $\, c_1 \geq \ldots \geq c_s \,$ of positive
integers giving the lengths of its $r$ rows and $s$ columns, respectively,%
\footnote{It is to be understood that $\, b_r > 0 \,$ and $\, c_s > 0 \,$
but $\, b_i = 0 \,$ if $\, i > r \,$ and $\, c_j = 0 \,$ if $\, j > s$.},
will be an allowed Young super\-diagram for $\mbox{\fraktw sl}(m\,|\,n)$
if and only if $\, b_{m+1} \leq n$; in this case it will describe an
irreducible representation of $\mbox{\fraktw sl}(m\,|\,n)$ whose
Kac-Dynkin labels $\, l_1,\ldots,l_{m+n-1} \,$ can be found as follows.
Define the reduced column lengths by
\begin{equation} \label{eq:KDLSL1}
  c'_j~=~(c_j - m) \, \theta(c_j - m)~,
\end{equation}
where $\theta$ is the step function; then
\begin{eqnarray} \label{eq:KDLSL2}
 &l_i~=~b_i - b_{i+1} \qquad \mbox{for}~i = 1,\ldots,m-1~,& \nonumber \\
 &l_m~=~b_m + c'_1~,& \\
 &l_{m+j}~=~c'_j - c'_{j+1} \qquad \mbox{for}~j = 1,\ldots,n-1~. \nonumber
\end{eqnarray}

On the other hand, the branching rules under reduction from the Lie
superalgebra $\mbox{\fraktw sl}(m\,|\,n)$ to the semisimple part
$\, \mbox{\fraktw sl}(m) \oplus \mbox{\fraktw sl}(n) \,$ of its even
subalgebra in terms of Young diagrams and superdiagrams can, according
to ref.\ \cite{BB}, be derived immediately from the corresponding
branching rules under reduction from the ordinary Lie algebra
$\mbox{\fraktw sl}(m+n)$ under restriction to the same subalgebra
$\, \mbox{\fraktw sl}(m) \oplus \mbox{\fraktw sl}(n)$, which in
turn are given in ref.\ \cite{IN}, for a large class of examples.
In fact all that needs to be done is to replace the Young diagram
for the second summand $\mbox{\fraktw sl}(n)$, which represents the
odd sector of the representation space, by its transposed diagram,
exchanging rows and columns. As an example, we show on the next
page the decomposition of the Young superdiagram given by
$\, r\!=\!2$, $s\!=\!2 \,$ with $\, b_1 = 3 = c_1$, $b_2 = 2
= c_2 \,$ and $\, b_3 = 1 = c_3 \,$ which, according to eqns
(\ref{eq:KDLSL1}) and (\ref{eq:KDLSL2}), corresponds to the
typical codon representation of $\mbox{\fraktw sl}(3\,|\,1)$,
of highest weight $(1,1,l_3)$ with $l_3\!=\!1$, as well as
to the typical codon representation of \linebreak
\begin{figure}[h]
\begin{minipage}[b]{14cm} 
\unitlength4mm
\begin{picture}(35,5)
 \linethickness{0.2mm}
 \multiput(0,4)(1,0){3}{\line(1,1){0.95}}
 \multiput(0,4)(1,0){3}{\framebox(0.95,0.95){}}
 \multiput(0,3)(1,0){2}{\line(1,1){0.95}}
 \multiput(0,3)(1,0){2}{\framebox(0.95,0.95){}}
 \put(0,2){\line(1,1){0.95}}
 \put(0,2){\framebox(0.95,0.95){}}

 \put(1.5,1.25){\makebox(0,0){{\small $\mbox{\frakte sl}(3\,|\,1)$}}}
 \put(1.5,0.25){\makebox(0,0){{\small $\mbox{\frakte sl}(2\,|\,2)$}}}

 \put(4,3.2){$\longrightarrow$}
 
 \put(6.9,3.2){$\Bigg($}
 \multiput(8,4)(1,0){3}{\framebox(0.95,0.95){}}
 \multiput(8,3)(1,0){2}{\framebox(0.95,0.95){}}
 \put(8,2){\framebox(0.95,0.95){}}
 \put(11.4,3.2){$, \, 1$}
 \put(12.6,3.2){$\Bigg)$}

 \put(10.1,1.25){\makebox(0,0){{\small $(1,1)$}}}
 \put(10.1,0.25){\makebox(0,0){{\small illegal}}}

 \put(13.7,3.2){+}

 \put(14.9,3.2){$\Bigg($}
 \put(15.6,3.2){$1 \; ,$}
 \multiput(17.2,4)(1,0){3}{\framebox(0.95,0.95){}}
 \multiput(17.2,3)(1,0){2}{\framebox(0.95,0.95){}}
 \put(17.2,2){\framebox(0.95,0.95){}}
 \put(20.6,3.2){$\Bigg)$}

 \put(18.1,1.25){\makebox(0,0){{\small illegal}}}
 \put(18.1,0.25){\makebox(0,0){{\small illegal}}}

 \put(21.7,3.2){$+~\, 2 \; \times$}

 \put(24.9,3.2){$\Bigg($}
 \multiput(26,4)(1,0){2}{\framebox(0.95,0.95){}}
 \put(26,3){\framebox(0.95,0.95){}}
 \put(28.4,3.2){$,$}
 \multiput(29.2,4)(1,0){2}{\framebox(0.95,0.95){}}
 \put(29.2,3){\framebox(0.95,0.95){}}
 \put(31.6,3.2){$\Bigg)$}

 \put(28.6,1.25){\makebox(0,0){{\small illegal}}}
 \put(28.6,0.25){\makebox(0,0){{\small $(1-1)$}}}
\end{picture}
\end{minipage}

\vspace{4mm}

\begin{minipage}[b]{14cm} 
\unitlength4mm
\begin{picture}(35,5)
 \linethickness{0.2mm}
 \put(6.7,3.2){+}
 
 \put(7.9,3.2){$\Bigg($}
 \multiput(9,4)(1,0){3}{\framebox(0.95,0.95){}}
 \multiput(9,3)(1,0){2}{\framebox(0.95,0.95){}}
 \put(12.4,3.2){$,$}
 \put(13.2,4){\framebox(0.95,0.95){}}
 \put(14.6,3.2){$\Bigg)$}

 \put(11.6,1.25){\makebox(0,0){{\small $(1,2)$}}}
 \put(11.6,0.25){\makebox(0,0){{\small $(1-1)$}}}

 \put(15.7,3.2){+}

 \put(16.9,3.2){$\Bigg($}
 \multiput(18,4)(1,0){3}{\framebox(0.95,0.95){}}
 \put(18,3){\framebox(0.95,0.95){}}
 \put(18,2){\framebox(0.95,0.95){}}
 \put(21.4,3.2){$,$}
 \put(22.2,4){\framebox(0.95,0.95){}}
 \put(23.6,3.2){$\Bigg)$}

 \put(20.6,1.25){\makebox(0,0){{\small $(2,0)$}}}
 \put(20.6,0.25){\makebox(0,0){{\small illegal}}}

 \put(24.7,3.2){+}

 \put(25.9,3.2){$\Bigg($}
 \multiput(27,4)(1,0){2}{\framebox(0.95,0.95){}}
 \multiput(27,3)(1,0){2}{\framebox(0.95,0.95){}}
 \put(27,2){\framebox(0.95,0.95){}}
 \put(29.4,3.2){$,$}
 \put(30.2,4){\framebox(0.95,0.95){}}
 \put(31.6,3.2){$\Bigg)$}

 \put(29.1,1.25){\makebox(0,0){{\small $(0,1)$}}}
 \put(29.1,0.25){\makebox(0,0){{\small illegal}}}
\end{picture}
\end{minipage}

\vspace{4mm}

\begin{minipage}[b]{14cm} 
\unitlength4mm
\begin{picture}(35,5)
 \linethickness{0.2mm}
 \put(6.7,3.2){+}
 
 \put(7.9,3.2){$\Bigg($}
 \multiput(9,4)(1,0){3}{\framebox(0.95,0.95){}}
 \put(9,3){\framebox(0.95,0.95){}}
 \put(12.4,3.2){$,$}
 \put(13.2,4){\framebox(0.95,0.95){}}
 \put(13.2,3){\framebox(0.95,0.95){}}
 \put(14.6,3.2){$\Bigg)$}

 \put(11.6,1.25){\makebox(0,0){{\small illegal}}}
 \put(11.6,0.25){\makebox(0,0){{\small $(2-0)$}}}

 \put(15.7,3.2){+}

 \put(16.9,3.2){$\Bigg($}
 \multiput(18,4)(1,0){2}{\framebox(0.95,0.95){}}
 \multiput(18,3)(1,0){2}{\framebox(0.95,0.95){}}
 \put(20.4,3.2){$,$}
 \put(21.2,4){\framebox(0.95,0.95){}}
 \put(21.2,3){\framebox(0.95,0.95){}}
 \put(22.6,3.2){$\Bigg)$}

 \put(20.1,1.25){\makebox(0,0){{\small illegal}}}
 \put(20.1,0.25){\makebox(0,0){{\small $(0-0)$}}}

 \put(23.7,3.2){+}

 \put(24.9,3.2){$\Bigg($}
 \multiput(26,4)(1,0){2}{\framebox(0.95,0.95){}}
 \put(26,3){\framebox(0.95,0.95){}}
 \put(26,2){\framebox(0.95,0.95){}}
 \put(28.4,3.2){$,$}
 \put(29.2,4){\framebox(0.95,0.95){}}
 \put(29.2,3){\framebox(0.95,0.95){}}
 \put(30.6,3.2){$\Bigg)$}

 \put(28.1,1.25){\makebox(0,0){{\small illegal}}}
 \put(28.1,0.25){\makebox(0,0){{\small illegal}}}
\end{picture}
\end{minipage}

\vspace{4mm}

\begin{minipage}[b]{14cm} 
\unitlength4mm
\begin{picture}(35,5)
 \linethickness{0.2mm}
 \put(6.7,3.2){+}
 
 \put(7.9,3.2){$\Bigg($}
 \multiput(9,4)(1,0){3}{\framebox(0.95,0.95){}}
 \put(9,3){\framebox(0.95,0.95){}}
 \put(12.4,3.2){$,$}
 \multiput(13.2,4)(1,0){2}{\framebox(0.95,0.95){}}
 \put(15.6,3.2){$\Bigg)$}

 \put(12.1,1.25){\makebox(0,0){{\small $(2,1)$}}}
 \put(12.1,0.25){\makebox(0,0){{\small $(2-2)$}}}

 \put(16.7,3.2){+}

 \put(17.9,3.2){$\Bigg($}
 \multiput(19,4)(1,0){2}{\framebox(0.95,0.95){}}
 \multiput(19,3)(1,0){2}{\framebox(0.95,0.95){}}
 \put(21.4,3.2){$,$}
 \multiput(22.2,4)(1,0){2}{\framebox(0.95,0.95){}}
 \put(24.6,3.2){$\Bigg)$}

 \put(21.6,1.25){\makebox(0,0){{\small $(0,2)$}}}
 \put(21.6,0.25){\makebox(0,0){{\small $(0-2)$}}}

 \put(25.7,3.2){+}

 \put(26.9,3.2){$\Bigg($}
 \multiput(28,4)(1,0){2}{\framebox(0.95,0.95){}}
 \put(28,3){\framebox(0.95,0.95){}}
 \put(28,2){\framebox(0.95,0.95){}}
 \put(30.4,3.2){$,$}
 \multiput(31.2,4)(1,0){2}{\framebox(0.95,0.95){}}
 \put(33.6,3.2){$\Bigg)$}

 \put(30.6,1.25){\makebox(0,0){{\small $(1,0)$}}}
 \put(30.6,0.25){\makebox(0,0){{\small illegal}}}
\end{picture}
\end{minipage}

\vspace{4mm}

\begin{minipage}[b]{14cm} 
\unitlength4mm
\begin{picture}(35,5)
 \linethickness{0.2mm}
 \put(6.7,3.2){+}
 
 \put(7.9,3.2){$\Bigg($}
 \multiput(9,4)(1,0){2}{\framebox(0.95,0.95){}}
 \put(9,3){\framebox(0.95,0.95){}}
 \put(11.4,3.2){$,$}
 \put(12.2,4){\framebox(0.95,0.95){}}
 \put(12.2,3){\framebox(0.95,0.95){}}
 \put(12.2,2){\framebox(0.95,0.95){}}
 \put(13.6,3.2){$\Bigg)$}

 \put(11.1,1.25){\makebox(0,0){{\small illegal}}}
 \put(11.1,0.25){\makebox(0,0){{\small illegal}}}

 \put(14.7,3.2){+}

 \put(15.9,3.2){$\Bigg($}
 \multiput(17,4)(1,0){3}{\framebox(0.95,0.95){}}
 \put(20.4,3.2){$,$}
 \multiput(21.2,4)(1,0){2}{\framebox(0.95,0.95){}}
 \put(21.2,3){\framebox(0.95,0.95){}}
 \put(23.6,3.2){$\Bigg)$}

 \put(19.6,1.25){\makebox(0,0){{\small illegal}}}
 \put(19.6,0.25){\makebox(0,0){{\small $(3-1)$}}}
\end{picture}
\end{minipage}

\vspace{4mm}

\begin{minipage}[b]{14cm} 
\unitlength4mm
\begin{picture}(35,5)
 \linethickness{0.2mm}
 \put(6.7,3.2){+}
 
 \put(7.9,3.2){$\Bigg($}
 \put(9,4){\framebox(0.95,0.95){}}
 \put(10.4,3.2){$,$}
 \multiput(11.2,4)(1,0){3}{\framebox(0.95,0.95){}}
 \multiput(11.2,3)(1,0){2}{\framebox(0.95,0.95){}}
 \put(14.6,3.2){$\Bigg)$}

 \put(11.6,1.25){\makebox(0,0){{\small illegal}}}
 \put(11.6,0.25){\makebox(0,0){{\small $(1-1)$}}}

 \put(15.7,3.2){+}

 \put(16.9,3.2){$\Bigg($}
 \put(18,4){\framebox(0.95,0.95){}}
 \put(19.4,3.2){$,$}
 \multiput(20.2,4)(1,0){3}{\framebox(0.95,0.95){}}
 \put(20.2,3){\framebox(0.95,0.95){}}
 \put(20.2,2){\framebox(0.95,0.95){}}
 \put(23.6,3.2){$\Bigg)$}

 \put(20.6,1.25){\makebox(0,0){{\small illegal}}}
 \put(20.6,0.25){\makebox(0,0){{\small illegal}}}

 \put(24.7,3.2){+}

 \put(25.9,3.2){$\Bigg($}
 \put(27,4){\framebox(0.95,0.95){}}
 \put(28.4,3.2){$,$}
 \multiput(29.2,4)(1,0){2}{\framebox(0.95,0.95){}}
 \multiput(29.2,3)(1,0){2}{\framebox(0.95,0.95){}}
 \put(29.2,2){\framebox(0.95,0.95){}}
 \put(31.6,3.2){$\Bigg)$}

 \put(29.1,1.25){\makebox(0,0){{\small illegal}}}
 \put(29.1,0.25){\makebox(0,0){{\small illegal}}}
\end{picture}
\end{minipage}

\vspace{4mm}

\begin{minipage}[b]{14cm} 
\unitlength4mm
\begin{picture}(35,5)
 \linethickness{0.2mm}
 \put(6.7,3.2){+}
 
 \put(7.9,3.2){$\Bigg($}
 \put(9,4){\framebox(0.95,0.95){}}
 \put(9,3){\framebox(0.95,0.95){}}
 \put(10.4,3.2){$,$}
 \multiput(11.2,4)(1,0){3}{\framebox(0.95,0.95){}}
 \put(11.2,3){\framebox(0.95,0.95){}}
 \put(14.6,3.2){$\Bigg)$}

 \put(11.6,1.25){\makebox(0,0){{\small illegal}}}
 \put(11.6,0.25){\makebox(0,0){{\small $(0-2)$}}}

 \put(15.7,3.2){+}

 \put(16.9,3.2){$\Bigg($}
 \put(18,4){\framebox(0.95,0.95){}}
 \put(18,3){\framebox(0.95,0.95){}}
 \put(19.4,3.2){$,$}
 \multiput(20.2,4)(1,0){2}{\framebox(0.95,0.95){}}
 \multiput(20.2,3)(1,0){2}{\framebox(0.95,0.95){}}
 \put(22.6,3.2){$\Bigg)$}

 \put(20.1,1.25){\makebox(0,0){{\small illegal}}}
 \put(20.1,0.25){\makebox(0,0){{\small $(0-0)$}}}

 \put(23.7,3.2){+}

 \put(24.9,3.2){$\Bigg($}
 \put(26,4){\framebox(0.95,0.95){}}
 \put(26,3){\framebox(0.95,0.95){}}
 \put(27.4,3.2){$,$}
 \multiput(28.2,4)(1,0){2}{\framebox(0.95,0.95){}}
 \put(28.2,3){\framebox(0.95,0.95){}}
 \put(28.2,2){\framebox(0.95,0.95){}}
 \put(30.6,3.2){$\Bigg)$}

 \put(28.1,1.25){\makebox(0,0){{\small illegal}}}
 \put(28.1,0.25){\makebox(0,0){{\small illegal}}}
\end{picture}
\end{minipage}

\vspace{4mm}

\begin{minipage}[b]{14cm} 
\unitlength4mm
\begin{picture}(35,5)
 \linethickness{0.2mm}
 \put(6.7,3.2){+}
 
 \put(7.9,3.2){$\Bigg($}
 \multiput(9,4)(1,0){2}{\framebox(0.95,0.95){}}
 \multiput(12.2,4)(1,0){3}{\framebox(0.95,0.95){}}
 \put(11.4,3.2){$,$}
 \put(12.2,3){\framebox(0.95,0.95){}}
 \put(15.6,3.2){$\Bigg)$}

 \put(12.1,1.25){\makebox(0,0){{\small illegal}}}
 \put(12.1,0.25){\makebox(0,0){{\small $(2-2)$}}}

 \put(16.7,3.2){+}

 \put(17.9,3.2){$\Bigg($}
 \multiput(19,4)(1,0){2}{\framebox(0.95,0.95){}}
 \put(21.4,3.2){$,$}
 \multiput(22.2,4)(1,0){2}{\framebox(0.95,0.95){}}
 \multiput(22.2,3)(1,0){2}{\framebox(0.95,0.95){}}
 \put(24.6,3.2){$\Bigg)$}

 \put(21.6,1.25){\makebox(0,0){{\small illegal}}}
 \put(21.6,0.25){\makebox(0,0){{\small $(2-0)$}}}

 \put(25.7,3.2){+}

 \put(26.9,3.2){$\Bigg($}
 \multiput(28,4)(1,0){2}{\framebox(0.95,0.95){}}
 \put(30.4,3.2){$,$}
 \multiput(31.2,4)(1,0){2}{\framebox(0.95,0.95){}}
 \put(31.2,3){\framebox(0.95,0.95){}}
 \put(31.2,2){\framebox(0.95,0.95){}}
 \put(33.6,3.2){$\Bigg)$}

 \put(30.6,1.25){\makebox(0,0){{\small illegal}}}
 \put(30.6,0.25){\makebox(0,0){{\small illegal}}}
\end{picture}
\end{minipage}

\vspace{4mm}

\begin{minipage}[b]{14cm} 
\unitlength4mm
\begin{picture}(35,5)
 \linethickness{0.2mm}
 \put(6.7,3.2){+}
 
 \put(7.9,3.2){$\Bigg($}
 \put(9,4){\framebox(0.95,0.95){}}
 \put(9,3){\framebox(0.95,0.95){}}
 \put(9,2){\framebox(0.95,0.95){}}
 \put(10.4,3.2){$,$}
 \multiput(11.2,4)(1,0){2}{\framebox(0.95,0.95){}}
 \put(11.2,3){\framebox(0.95,0.95){}}
 \put(13.6,3.2){$\Bigg)$}

 \put(11.1,1.25){\makebox(0,0){{\small illegal}}}
 \put(11.1,0.25){\makebox(0,0){{\small illegal}}}

 \put(14.7,3.2){+}

 \put(15.9,3.2){$\Bigg($}
 \multiput(17,4)(1,0){2}{\framebox(0.95,0.95){}}
 \put(17,3){\framebox(0.95,0.95){}}
 \put(19.4,3.2){$,$}
 \multiput(20.2,4)(1,0){3}{\framebox(0.95,0.95){}}
 \put(23.6,3.2){$\Bigg)$}

 \put(20.1,1.25){\makebox(0,0){{\small $(1,1)$}}}
 \put(20.1,0.25){\makebox(0,0){{\small $(1-3)$}}}
\end{picture}
\end{minipage}
\end{figure}

\clearpage

\noindent
$\mbox{\fraktw sl}(2\,|\,2)$, of highest weight $(1,l_2,1)$ with
$l_2\!=\!3$. The highest weights with respect to $\mbox{\fraktw sl}(3)$
and to $\, \mbox{\fraktw sl}(2) \oplus \mbox{\fraktw sl}(2) \,$ corresponding
to the ordinary Young diagrams resulting from this decomposition are also
exhibited and the ``illegal'' diagrams are identified: they are the ones
that must be eliminated to comply with the prescription that Young diagrams
for $\mbox{\fraktw sl}(k)$ must not have more than $k$ rows. In this way,
we arrive at the branching schemes for the typical codon representations
of $\mbox{\fraktw sl}(m\,|\,n)$ given in Table 1 and in Table 2, since
the remaining cases can be checked directly from the rules given in
Table 1 of ref.\ \cite{BB}.

For the orthosymplectic Lie superalgebras $\mbox{\fraktw osp}(M\,|\,N)$,
where $\, M = 2m+1$ \linebreak or $\, M = 2m \,$ and $\, N = 2n$, the
procedure is somewhat more complicated; it is described in ref.\ \cite{MSS}.
First of all, it must be noted that the construction and interpretation of
Young super\-diagrams for $\mbox{\fraktw osp}(M\,|\,N)$, as compared to that
for $\mbox{\fraktw sl}(m\,|\,n)$, is subject to the same adjustments as that of
ordinary Young diagrams for $\mbox{\fraktw sp}(N)$ and $\mbox{\fraktw so}(M)$,
as compared to that for $\mbox{\fraktw sl}(n)$: in particular, they may
contain ``spinor'' or ``half'' boxes (referring, of course, only to the
$\mbox{\fraktw so}(M)$ part of the even subalgebra) which by convention
will be located in the $(n+1)^{\rm st}$ row. \linebreak We shall follow the
notation of ref.\ \cite{MSS}, except that we shall continue to distinguish
Young superdiagrams from their non-supersymmetric counterparts by the
insertion of a diagonal line across each box, including the ``spinor''
or ``half'' boxes. The relation between the lengths $\, b_1 \geq \ldots
\geq b_n \,$ of the first $n$ rows and $\, c_1 \geq \ldots \geq c_m \,$
of the first $m$ columns on the one hand and the Kac-Dynkin labels
$\, l_1,\ldots, l_{n-1},l_n,l_{n+1},\ldots,l_{n+m} \,$ on the other
hand is summarized in eqns (3.1), (3.4) and (3.5) of ref.\ \cite{MSS}.
The prescription for determining the branching rules under reduction
from the Lie superalgebra $\mbox{\fraktw osp}(M\,|\,N)$ to its even
part $\, \mbox{\fraktw sp}(N) \oplus \mbox{\fraktw so}(M) \,$ has also
been determined and is formally summarized in eqns (3.2), (3.3) and
(3.6) of ref.\ \cite{MSS}. The starting point is to dissect the given
Young super\-diagram into two ordinary Young diagrams: one for the
$\mbox{\fraktw sp}(N)$ part formed by the first $n$ rows and one
for the $\mbox{\fraktw so}(M)$ part formed by the remaining rows,
but reflected along the main diagonal. Together, they stand for
the irreducible subrepresentation of the even subalgebra
$\, \mbox{\fraktw sp}(N) \oplus \mbox{\fraktw so}(M) \,$
generated from the original highest weight vector by application
of all even generators. It forms the ground floor of a building
in which all the other irreducible subrepresentations of the even
subalgebra are arranged in higher floors, each counted according to
the minimum number of odd generators required to reach it from the
ground floor. The procedure for determining which Young diagrams
describe the irreducible subrepresentations that do appear in the
higher floors is complicated, requiring the use of generalized Young
diagrams for $\mbox{\fraktw sp}(N)$ with negative boxes, as introduced
in ref.\ \cite{GSS2}, that must be multiplied to standard Young diagrams for
$\mbox{\fraktw so}(M)$, plus rules for eliminating Young diagrams resulting
from this process that represent non-tracefree parts. A discussion of the
general formulas presented in ref.\ \cite{MSS} is not very instructive, so
we prefer to just illustrate them by presenting two important examples:
\linebreak

\pagebreak

\noindent
the branching schemes for the typical codon representations of
$\mbox{\fraktw osp}(4\,|\,2)$ with highest weight $({7 \over 2},0,1)$
and of $\mbox{\fraktw osp}(5\,|\,2)$ with highest weight $({5 \over 2},0,1)$.

We begin by calculating the $\mbox{\fraktw sp}(2) \oplus \mbox{\fraktw so}(4)$
content of the typical codon representation of $\mbox{\fraktw osp}(4\,|\,2)$
with highest weight $({7 \over 2},0,1)$. According to eqn (3.4) of ref.\
\cite{MSS}, the labels $\, b_1 \geq \ldots \geq b_n \,$ and $\, c_1 \geq
\ldots \geq c_m \,$ of the corresponding Young superdiagram are given by
\begin{eqnarray}
 &b_1~=~l^0_1~=~l_1 - {\textstyle {1 \over 2}} (l_2 + l_3)~=~3~,& \nonumber \\
 &c_1~=~n + {\textstyle {1 \over 2}} (l_3 + l_2)~
      =~    {\textstyle {3 \over 2}}~,& \\
 &c_2~=~n + {\textstyle {1 \over 2}} (l_3 - l_2)~
      =~    {\textstyle {3 \over 2}}~,& \nonumber
\end{eqnarray}
so the Young superdiagram has the form indicated in Table 3:
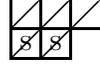
\begin{center}
\unitlength4mm
\begin{picture}(3,2)
\linethickness{0.2mm}
 \multiput(0,1)(1,0){3}{\line(1,1){0.95}}
 \multiput(0,1)(1,0){3}{\framebox(0.95,0.95){}}
 \multiput(0,0)(1,0){2}{\line(1,1){0.95}}
 \multiput(0,0)(1,0){2}{\framebox(0.95,0.95){s}}
\end{picture}
\vspace{-2mm}
\end{center}
Therefore, the Young diagram for the even subalgebra $\, \mbox{\fraktw sp}(2)
\oplus \mbox{\fraktw so}(4) \,$ is:
\begin{center}
\unitlength4mm
\begin{picture}(7,2)
\linethickness{0.2mm}
 \put(0.1,0.7){$\Bigg($}
 \multiput(1.2,1)(1,0){3}{\framebox(0.95,0.95){}}
 \put(4.6,0.7){$,$}
 \put(5.4,1){\framebox(0.95,0.95){s}}
 \put(5.4,0){\framebox(0.95,0.95){s}}
 \put(6.8,0.7){$\Bigg)$}
\end{picture}
\vspace{-2mm}
\end{center}
It describes the irreducible representation of highest weight $\, (3)-(0,1) \,$
which forms the ground floor. The irreducible representations on the following
floors are computed graphically as follows: \\[7mm]
1. floor: \hspace{\fill}
\begin{minipage}{12.4cm}
\unitlength4mm
\begin{picture}(31,2)
\linethickness{0.2mm}
 \put(0.1,0.7){$\Bigg($}
 \put(1.2,0){\framebox(0.95,0.95){}}
 \put(2.18,1){\line(0,1){1}}
 \put(2.8,0.7){$\times$}
 \multiput(4.2,1)(1,0){3}{\framebox(0.95,0.95){}}
 \put(7.6,0.7){$,$}
 \put(8.4,1){\framebox(0.95,0.95){}}
 \put(10,0.7){$\times$}
 \put(11.4,1){\framebox(0.95,0.95){s}}
 \put(11.4,0){\framebox(0.95,0.95){s}}
 \put(12.8,0.7){$\Bigg)$}
 \put(13.9,0.7){=}
 \put(15.1,0.7){$\Bigg($}
 \multiput(16.2,1)(1,0){2}{\framebox(0.95,0.95){}}
 \put(18.6,0.7){$,$}
 \put(19.4,1){\framebox(0.95,0.95){s}}
 \put(19.4,0){\framebox(0.95,0.95){s}}
 \put(20.4,1){\framebox(0.95,0.95){}}
 \put(22,0.7){+}
 \put(24.4,1){\framebox(0.95,0.95){s}}
 \put(23.4,0){\framebox(0.95,0.95){s}}
 \put(25.8,0.7){$\Bigg)$,}
\end{picture}
\end{minipage} \\[7mm]
corresponding to the highest weights $\, (2)-(1,2) \,$ and $\, (2)-(1,0)$,
\\[7mm]
2. floor: \hspace{\fill}
\begin{minipage}{12.4cm}
\unitlength4mm
\begin{picture}(31,2)
\linethickness{0.2mm}
 \put(0.1,0.7){$\Bigg($}
 \multiput(1.2,0)(1,0){2}{\framebox(0.95,0.95){}}
 \put(3.16,1){\line(0,1){1}}
 \put(3.8,0.7){$\times$}
 \multiput(5.2,1)(1,0){3}{\framebox(0.95,0.95){}}
 \put(8.6,0.7){$,$}
 \put(9.4,1){\framebox(0.95,0.95){}}
 \put(9.4,0){\framebox(0.95,0.95){}}
 \put(11,0.7){$\times$}
 \put(12.4,1){\framebox(0.95,0.95){s}}
 \put(12.4,0){\framebox(0.95,0.95){s}}
 \put(13.8,0.7){$\Bigg)$}
 \put(14.9,0.7){=}
 \put(16.1,0.7){$\Bigg($}
 \put(17.2,1){\framebox(0.95,0.95){}}
 \put(18.6,0.7){$,$}
 \put(19.4,1){\framebox(0.95,0.95){s}}
 \put(19.4,0){\framebox(0.95,0.95){s}}
 \put(20.4,1){\framebox(0.95,0.95){}}
 \put(20.4,0){\framebox(0.95,0.95){}}
 \put(22,0.7){+}
 \put(23.4,0){\framebox(0.95,0.95){s}}
 \put(24.4,1){\framebox(0.95,0.95){s}}
 \put(25.4,1){\framebox(0.95,0.95){}}
 \put(27,0.7){+}
 \put(28.4,0){\framebox(0.95,0.95){s}}
 \put(28.4,1){\framebox(0.95,0.95){s}}
 \put(29.8,0.7){$\Bigg)$,}
\end{picture}
\end{minipage} \\[7mm]
corresponding to the highest weights $\, (1)-(0,3)$, $(1)-(2,1) \,$ and
$\, (1)-(0,1)$, \\[7mm]
3. floor: \hspace{\fill}
\begin{minipage}{12.4cm}
\unitlength4mm
\begin{picture}(31,2)
\linethickness{0.2mm}
 \put(0.1,0.7){$\Bigg($}
 \multiput(1.2,0)(1,0){3}{\framebox(0.95,0.95){}}
 \put(4.14,1){\line(0,1){1}}
 \put(4.8,0.7){$\times$}
 \multiput(6.2,1)(1,0){3}{\framebox(0.95,0.95){}}
 \put(9.6,0.7){$,$}
 \put(10.4,1){\framebox(0.95,0.95){}}
 \put(12,0.7){$\times$}
 \put(13.4,1){\framebox(0.95,0.95){s}}
 \put(13.4,0){\framebox(0.95,0.95){s}}
 \put(14.8,0.7){$\Bigg)$}
 \put(15.9,0.7){=}
 \put(17.1,0.7){$\Bigg($}
 \put(17.7,0.7){$1 \; ,$}
 \put(19.4,1){\framebox(0.95,0.95){s}}
 \put(19.4,0){\framebox(0.95,0.95){s}}
 \put(20.4,1){\framebox(0.95,0.95){}}
 \put(22,0.7){+}
 \put(23.4,0){\framebox(0.95,0.95){s}}
 \put(24.4,1){\framebox(0.95,0.95){s}}
 \put(25.8,0.7){$\Bigg)$,}
\end{picture}
\end{minipage} \\[7mm]
corresponding to the highest weights $\, (0)-(1,2) \,$ and $\, (0)-(1,0)$.
These are precisely the highest weights listed in Table 3 for this case.

\pagebreak

We proceed to calculate the $\mbox{\fraktw sp}(2) \oplus \mbox{\fraktw so}(5)$
content of the typical codon representation of $\mbox{\fraktw osp}(5\,|\,2)$
with highest weight $({5 \over 2},0,1)$. According to eqn (3.1) of ref.\
\cite{MSS}, the labels $\, b_1 \geq \ldots \geq b_n \,$ and $\, c_1 \geq
\ldots \geq c_m \,$ of the corresponding Young superdiagram are given by
\begin{eqnarray}
 &b_1~=~l^0_1~=~l_1 - l_2 - {\textstyle {1 \over 2}} l_3~=~2~,& \nonumber \\
 &c_1~=~n + l_2 + {\textstyle {1 \over 2}} l_3~
      =~  {\textstyle {3 \over 2}}~,& \\
 &c_2~=~n + {\textstyle {1 \over 2}} l_3~
      =~  {\textstyle {3 \over 2}}~,& \nonumber
\end{eqnarray}
so the Young superdiagram has the form indicated in Table 3:
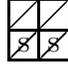
\begin{center}
\unitlength4mm
\begin{picture}(2,2)
\linethickness{0.2mm}
 \multiput(0,1)(1,0){2}{\line(1,1){0.95}}
 \multiput(0,1)(1,0){2}{\framebox(0.95,0.95){}}
 \multiput(0,0)(1,0){2}{\line(1,1){0.95}}
 \multiput(0,0)(1,0){2}{\framebox(0.95,0.95){s}}
\end{picture}
\vspace{-2mm}
\end{center}
Therefore, the Young diagram for the even subalgebra $\, \mbox{\fraktw sp}(2)
\oplus \mbox{\fraktw so}(5) \,$ is:
\begin{center}
\unitlength4mm
\begin{picture}(6,2)
\linethickness{0.2mm}
 \put(0.1,0.7){$\Bigg($}
 \multiput(1.2,1)(1,0){2}{\framebox(0.95,0.95){}}
 \put(3.6,0.2){$,$}
 \put(4.4,1){\framebox(0.95,0.95){s}}
 \put(4.4,0){\framebox(0.95,0.95){s}}
 \put(5.8,0.7){$\Bigg)$}
\end{picture}
\vspace{-2mm}
\end{center}
It describes the irreducible representation of highest weight $\, (2)-(0,1) \,$
which forms the ground floor. The irreducible representations on the following
floors are computed graphically as follows: \\[7mm]
1. floor: \hspace{\fill}
\begin{minipage}{11cm}
\unitlength4mm
\begin{picture}(27,2)
\linethickness{0.2mm}
 \put(0.1,0.7){$\Bigg($}
 \put(1.2,0){\framebox(0.95,0.95){}}
 \put(2.18,1){\line(0,1){1}}
 \put(2.8,0.7){$\times$}
 \multiput(4.2,1)(1,0){2}{\framebox(0.95,0.95){}}
 \put(6.6,0.7){$,$}
 \put(7.4,1){\framebox(0.95,0.95){}}
 \put(9,0.7){$\times$}
 \put(10.4,1){\framebox(0.95,0.95){s}}
 \put(10.4,0){\framebox(0.95,0.95){s}}
 \put(11.8,0.7){$\Bigg)$}
 \put(12.9,0.7){=}
 \put(14.1,0.7){$\Bigg($}
 \put(15.2,1){\framebox(0.95,0.95){}}
 \put(16.6,0.7){$,$}
 \put(17.4,1){\framebox(0.95,0.95){s}}
 \put(17.4,0){\framebox(0.95,0.95){s}}
 \put(18.4,1){\framebox(0.95,0.95){}}
 \put(19.8,0.7){$\Bigg)$,}
\end{picture}
\end{minipage} \\[7mm]
corresponding to the highest weight $\, (1)-(1,1)$, \\[7mm]
2. floor: \hspace{\fill}
\begin{minipage}{11cm}
\unitlength4mm
\begin{picture}(27,2)
\linethickness{0.2mm}
 \put(0.1,0.7){$\Bigg($}
 \multiput(1.2,0)(1,0){2}{\framebox(0.95,0.95){}}
 \put(3.16,1){\line(0,1){1}}
 \put(3.8,0.7){$\times$}
 \multiput(5.2,1)(1,0){2}{\framebox(0.95,0.95){}}
 \put(7.6,0.7){$,$}
 \put(8.4,1){\framebox(0.95,0.95){}}
 \put(8.4,0){\framebox(0.95,0.95){}}
 \put(10,0.7){$\times$}
 \put(11.4,1){\framebox(0.95,0.95){s}}
 \put(11.4,0){\framebox(0.95,0.95){s}}
 \put(12.8,0.7){$\Bigg)$}
 \put(13.9,0.7){=}
 \put(15.1,0.7){$\Bigg($}
 \put(15.7,0.7){$1 \; ,$}
 \put(17.4,1){\framebox(0.95,0.95){s}}
 \put(17.4,0){\framebox(0.95,0.95){s}}
 \put(18.4,1){\framebox(0.95,0.95){}}
 \put(18.4,0){\framebox(0.95,0.95){}}
 \put(19.8,0.7){$\Bigg)$,}
\end{picture}
\end{minipage} \\[7mm]
corresponding to the highest weight $\, (0)-(0,3)$. Again, these
are precisely the highest weights listed in Table 3 for this case.

Finally, it should be mentioned that we have omitted from Tables 1-3 some
of the typical codon representations determined in \cite{FS} because their
branching schemes are obvious from those that are listed. Examples are the
typical codon representations of $\mbox{\fraktw sl}(4\,|\,1)$ with highest
weight $(0,0,1,l_4)$ and of $\mbox{\fraktw sl}(2\,|\,2)$ with highest weight
$(0,l_2,3)$, which are complex conjugate to those with highest weight
$(1,0,0,l_4)$ and $(3,l_2,0)$, respectively, and which therefore exhibit
the same branching schemes, in all phases, except for complex conjugation
which however does not affect dimensions. Similarly, it is known that
the branching rules of typical representations of the Lie superalgebra
$\mbox{\fraktw osp}(4\,|\,2,\alpha)$ upon reduction to its even part
do not depend on $\alpha$ \cite{VJ}, so that we may without loss of
generality put $\, \alpha = 1$. Moreover, our calculations have shown
that the three typical representations with highest weight $(5,0,0)$,
$({7 \over 2},3,0)$ and $({7 \over 2},0,3)$, as well as the three
typical representations with highest weight $(3,1,1)$, $({7 \over 2},1,0)$
and $({7 \over 2},0,1)$, although inequivalent, have the same branching
rules under this reduction, so we have listed only one of each.

\begin{table}[ht] \label{tb:BRT11}
 \begin{center}
  Table 1: Branching of codon representations of type~I Lie superalgebras \\
           in the first step $\; \cg \smsupset \cg_{\bar{0}}^{\rm ss} \,$:
           Part 1
 \end{center}
 \begin{center}
 \begin{tabular}{|c|c|c|c|c|} \hline
  \rule{0ex}{3ex}            LSA            \rule{0ex}{3ex} &
    Highest weight   &    Young     &           Highest Weights              &
  $d$ \\
  \rule[-1.5ex]{0ex}{1.5ex} $\cg$ \rule[-1.5ex]{0ex}{1.5ex} &
  of $\cg$-multiplet & Superdiagram & of $\cg_{\bar{0}}^{\rm ss}$-multiplets &
  \\ \hline\hline
  \rule{0ex}{3ex} $\mbox{\fraktw sl}(2\,|\,1)$ & $(15,l_2)$
                    & $l_2 = 1$ &     $(16)$      &      17       \\
                  & & \unitlength5mm
                      \begin{picture}(8,0)
                      \linethickness{0.2mm}
                       \put(0,-1){\line(1,1){0.95}}
                       \put(0,-1){\framebox(0.95,0.95){}}
                       \put(1,-1){\framebox(5.95,0.95)
                                  {\ldots {\small\sf 14 boxes} \ldots}}
                       \put(7,-1){\line(1,1){0.95}}
                       \put(7,-1){\framebox(0.95,0.95){}}
                       \put(0,-2){\line(1,1){0.95}}
                       \put(0,-2){\framebox(0.95,0.95){}}
                      \end{picture}
                                & $2 \times (15)$ & $2 \times 16$ \\
                  & &           &     $(14)$      &      15
                  \\[3ex] \hline
  \rule{0ex}{3ex} $\mbox{\fraktw sl}(3\,|\,1)$ & $(1,1,l_3)$
                    & $l_3 = 1$ &     $(2,1)$      &      15      \\
                  & & \unitlength5mm
                      \begin{picture}(3,0)
                      \linethickness{0.2mm}
                       \multiput(0,-1)(1,0){3}{\line(1,1){0.95}}
                       \multiput(0,-1)(1,0){3}{\framebox(0.95,0.95){}}
                       \multiput(0,-2)(1,0){2}{\line(1,1){0.95}}
                       \multiput(0,-2)(1,0){2}{\framebox(0.95,0.95){}}
                       \put(0,-3){\line(1,1){0.95}}
                       \put(0,-3){\framebox(0.95,0.95){}}
                      \end{picture}
                                &     $(1,2)$      &      15      \\
                  & &           & $2 \times (1,1)$ & $2 \times 8$ \\
                  & &           &     $(2,0)$      &       6      \\
                  & &           &     $(0,2)$      &       6      \\
                  & &           &     $(1,0)$      &       3      \\
                  & &           &     $(0,1)$      &       3      \\[1ex] \hline
  \rule{0ex}{3ex} $\mbox{\fraktw sl}(4\,|\,1)$ & $(1,0,0,l_4)$
                    & $l_4 = 1$ &     $(1,1,0)$      &      20      \\
                  & & \unitlength5mm
                      \begin{picture}(2,0)
                      \linethickness{0.2mm}
                       \multiput(0,-1)(1,0){2}{\line(1,1){0.95}}
                       \multiput(0,-1)(1,0){2}{\framebox(0.95,0.95){}}
                       \put(0,-2){\line(1,1){0.95}}
                       \put(0,-2){\framebox(0.95,0.95){}}
                       \put(0,-3){\line(1,1){0.95}}
                       \put(0,-3){\framebox(0.95,0.95){}}
                       \put(0,-4){\line(1,1){0.95}}
                       \put(0,-4){\framebox(0.95,0.95){}}
                      \end{picture}
                                &     $(1,0,1)$      &      15      \\
                  & &           &     $(2,0,0)$      &      10      \\
                  & &           &     $(0,1,0)$      &       6      \\
                  & &           &     $(0,0,1)$      &       4      \\
                  & &           & $2 \times (1,0,0)$ & $2 \times 4$ \\
                  & &           &     $(0,0,0)$      &       1
                  \\[1ex] \hline
  \rule{0ex}{3ex} $\mbox{\fraktw sl}(6\,|\,1)$ & $(0,0,0,0,0,l_6)$
                    & $l_6 = 1$ &     $(0,0,1,0,0)$      &      20      \\
                  & & \unitlength5mm
                      \begin{picture}(1,0)
                      \linethickness{0.2mm}
                       \multiput(0,-6)(0,1){6}{\line(1,1){0.95}}
                       \multiput(0,-6)(0,1){6}{\framebox(0.95,0.95){}}
                      \end{picture}
                                &     $(0,1,0,0,0)$      &      15      \\
                  & &           &     $(0,0,0,1,0)$      &      15      \\
                  & &           &     $(1,0,0,0,0)$      &       6      \\
                  & &           &     $(0,0,0,0,1)$      &       6      \\
                  & &           & $2 \times (0,0,0,0,0)$ & $2 \times 1$
                  \\[6ex] \hline
 \end{tabular}
 \end{center}
\end{table}

\begin{table}[ht] \label{tb:BRT12}
 \begin{center}
  Table 2: Branching of codon representations of type~I Lie superalgebras \\
           in the first step $\; \cg \smsupset \cg_{\bar{0}}^{\rm ss} \,$:
           Part 2
 \end{center}
 \begin{center}
 \begin{tabular}{|c|c|c|c|c|} \hline
  \rule{0ex}{3ex}            LSA            \rule{0ex}{3ex} &
    Highest weight   &    Young     &           Highest Weights              &
  $d$ \\
  \rule[-1.5ex]{0ex}{1.5ex} $\cg$ \rule[-1.5ex]{0ex}{1.5ex} &
  of $\cg$-multiplet & Superdiagram & of $\cg_{\bar{0}}^{\rm ss}$-multiplets &
  \\ \hline\hline
  \rule{0ex}{3ex} $\mbox{\fraktw sl}(2\,|\,2)$ & $(3,l_2,0)$
                    & $l_2 = 2$ &       $(3) - (2)$      &       12      \\
                  & & \unitlength5mm
                      \begin{picture}(5,0)
                      \linethickness{0.2mm}
                       \multiput(0,-1)(1,0){5}{\line(1,1){0.95}}
                       \multiput(0,-1)(1,0){5}{\framebox(0.95,0.95){}}
                       \multiput(0,-2)(1,0){2}{\line(1,1){0.95}}
                       \multiput(0,-2)(1,0){2}{\framebox(0.95,0.95){}}
                      \end{picture}
                                & $2 \times ((4) - (1))$ & $2 \times 10$ \\
                  & &           &       $(5) - (0)$      &        6      \\
                  & &           & $2 \times ((2) - (1))$ & $2 \times 6$  \\
                  & &           & $3 \times ((3) - (0))$ & $3 \times 4$  \\
                  & &           &       $(1) - (0)$      &        2
                  \\[1ex] \cline{2-5}
  \rule{0ex}{3ex} & $(1,l_2,1)$
                    & $l_2 = 3$ & $2 \times ((2) - (2))$ & $2 \times 9$ \\
                  & & \unitlength5mm
                      \begin{picture}(3,0)
                      \linethickness{0.2mm}
                       \multiput(0,-1)(1,0){3}{\line(1,1){0.95}}
                       \multiput(0,-1)(1,0){3}{\framebox(0.95,0.95){}}
                       \multiput(0,-2)(1,0){2}{\line(1,1){0.95}}
                       \multiput(0,-2)(1,0){2}{\framebox(0.95,0.95){}}
                       \put(0,-3){\line(1,1){0.95}}
                       \put(0,-3){\framebox(0.95,0.95){}}
                      \end{picture}
                                &       $(3) - (1)$      &        8     \\
                  & &           &       $(1) - (3)$      &        8     \\
                  & &           & $4 \times ((1) - (1))$ & $4 \times 4$ \\
                  & &           & $2 \times ((2) - (0))$ & $2 \times 3$  \\
                  & &           & $2 \times ((0) - (2))$ & $2 \times 3$  \\
                  & &           & $2 \times ((0) - (0))$ & $2 \times 1$
                  \\[1ex] \hline
  \rule{0ex}{3ex} $\mbox{\fraktw sl}(3\,|\,2)$ & $(0,0,l_3,0)$
                    & $l_3 = 2$ &      $(1,1) - (1)$       &      16      \\
                  & & \unitlength5mm
                      \begin{picture}(2,0)
                      \linethickness{0.2mm}
                       \multiput(0,-3)(0,1){3}{\line(1,1){0.95}}
                       \multiput(0,-3)(0,1){3}{\framebox(0.95,0.95){}}
                       \multiput(1,-3)(0,1){3}{\line(1,1){0.95}}
                       \multiput(1,-3)(0,1){3}{\framebox(0.95,0.95){}}
                      \end{picture}
                                &      $(1,0) - (2)$       &       9      \\
                  & &           &      $(0,1) - (2)$       &       9      \\
                  & &           &      $(2,0) - (0)$       &       6      \\
                  & &           &      $(0,2) - (0)$       &       6      \\
                  & &           &      $(1,0) - (1)$       &       6      \\
                  & &           &      $(0,1) - (1)$       &       6      \\
                  & &           &      $(0,0) - (3)$       &       4      \\
                  & &           & $2 \times ((0,0) - (0))$ & $2 \times 1$
                  \\[1ex] \hline
  \rule{0ex}{3ex} $\mbox{\fraktw osp}(2\,|\,4)$ & $(l_1,1,0)$
                    & $l_1 = 1$ &     $(1,1)$      &      16       \\
                  & & \unitlength5mm
                      \begin{picture}(1,0)
                      \linethickness{0.2mm}
                       \put(0,-1){\line(1,1){0.95}}
                       \put(0,-1){\framebox(0.95,0.95){}}
                       \put(0,-2){\line(1,1){0.95}}
                       \put(0,-2){\framebox(0.95,0.95){}}
                      \end{picture}
                                & $2 \times (2,0)$ & $2 \times 10$ \\
                  & &           & $2 \times (0,1)$ & $2 \times 5$  \\
                  & &           & $4 \times (1,0)$ & $4 \times 4$  \\
                  & &           & $2 \times (0,0)$ & $2 \times 1$
                  \\[1ex] \hline
  \rule{0ex}{3ex} $\mbox{\fraktw osp}(2\,|\,6)$ & $(l_1,0,0,0)$
                    & $l_1 = 3$ &     $(0,0,1)$      &       14      \\
                  & & \unitlength5mm
                      \begin{picture}(2,0)
                      \linethickness{0.2mm}
                       \multiput(0,-1)(1,0){2}{\line(1,1){0.95}}
                       \multiput(0,-1)(1,0){2}{\framebox(0.95,0.95){}}
                      \end{picture}
                                & $2 \times (0,1,0)$ & $2 \times 14$ \\
                  & &           & $3 \times (1,0,0)$ & $3 \times 6$  \\
                  & &           & $4 \times (0,0,0)$ & $4 \times 1$
                  \\[1ex] \hline
 \end{tabular}
 \end{center}
\end{table}

\begin{table}[ht] \label{tb:BRT2}
 \begin{center}
  Table 3: Branching of codon representations of type~II Lie superalgebras \\
           in the first step $\; \cg \smsupset \cg_{\bar{0}} \,$
 \end{center}
 \begin{center}
 \begin{tabular}{|c|c|c|c|c|} \hline
  \rule{0ex}{3ex}            LSA            \rule{0ex}{3ex} &
    Highest weight   &    Young     &        Highest Weights        &
  $d$ \\
  \rule[-1.5ex]{0ex}{1.5ex} $\cg$ \rule[-1.5ex]{0ex}{1.5ex} &
  of $\cg$-multiplet & Superdiagram & of $\cg_{\bar{0}}$-multiplets &
  \\ \hline\hline
  \rule{0ex}{3ex} $\mbox{\fraktw osp}(3\,|\,2)$ & $({17 \over 2},15)$
                    & \unitlength5mm
                      \begin{picture}(1,0)
                      \linethickness{0.2mm}
                       \multiput(0,-9)(0,1){9}{\line(1,1){0.95}}
                       \multiput(0,-9)(0,1){9}{\framebox(0.95,0.95){}}
                       \put(0,-2){\makebox(0.95,0.95){s}}
                      \end{picture}
                      & $(1) - (15)$ & 32 \\
                  & & & $(0) - (17)$ & 18 \\
                  & & & $(0) - (13)$ & 14 \\[3.7cm] \hline
  \rule{0ex}{3ex} $\mbox{\fraktw osp}(3\,|\,4)$ & $(0,{5 \over 2},3)$
                    & \unitlength5mm
                      \begin{picture}(1,0)
                      \linethickness{0.2mm}
                       \multiput(0,-4)(0,1){4}{\framebox(0.95,0.95){}}
                       \multiput(0,-4)(0,1){4}{\line(1,1){0.95}}
                       \put(0,-3){\makebox(0.95,0.95){s}}
                      \end{picture}
                      & $(1,0) - (5)$ & 24 \\
                  & & & $(0,1) - (3)$ & 20 \\
                  & & & $(1,0) - (1)$ &  8 \\
                  & & & $(0,0) - (7)$ &  8 \\
                  & & & $(0,0) - (3)$ &  4 \\[1ex] \hline
  \rule{0ex}{3ex} $\mbox{\fraktw osp}(5\,|\,2)$ & $({5 \over 2},0,1)$
                    & \unitlength5mm
                      \begin{picture}(2,0)
                      \linethickness{0.2mm}
                       \multiput(0,-1)(1,0){2}{\line(1,1){0.95}}
                       \multiput(0,-1)(1,0){2}{\framebox(0.95,0.95){}}
                       \multiput(0,-2)(1,0){2}{\line(1,1){0.95}}
                       \multiput(0,-2)(1,0){2}{\framebox(0.95,0.95){}}
                       \put(0,-2){\framebox(0.95,0.95){s}}
                       \put(1,-2){\framebox(0.95,0.95){s}}
                      \end{picture}
                      & $(1) - (1,1)$ & 32 \\
                  & & & $(0) - (0,3)$ & 20 \\
                  & & & $(2) - (0,1)$ & 12 \\[1ex] \hline
  \rule{0ex}{3ex} $\mbox{\fraktw osp}(4\,|\,2)$ & $(5,0,0)$
                    & \unitlength5mm
                      \begin{picture}(5,0)
                      \linethickness{0.2mm}
                       \multiput(0,-1)(1,0){5}{\framebox(0.95,0.95){}}
                       \multiput(0,-1)(1,0){5}{\line(1,1){0.95}}
                      \end{picture}
                      & $(4) - (1) - (1)$ & 20 \\
                  & & & $(3) - (2) - (0)$ & 12 \\
                  & & & $(3) - (0) - (2)$ & 12 \\
                  & & & $(2) - (1) - (1)$ & 12 \\
                  & & & $(5) - (0) - (0)$ &  6 \\
                  & & & $(1) - (0) - (0)$ &  2 \\[1ex] \cline{2-5}
  \rule{0ex}{3ex} & $(\frac{7}{2},0,1)$
                    & \unitlength5mm
                      \begin{picture}(3,0)
                      \linethickness{0.2mm}
                       \multiput(0,-1)(1,0){3}{\line(1,1){0.95}}
                       \multiput(0,-1)(1,0){3}{\framebox(0.95,0.95){}}
                       \multiput(0,-2)(1,0){2}{\line(1,1){0.95}}
                       \multiput(0,-2)(1,0){2}{\framebox(0.95,0.95){s}}
                      \end{picture}
                      & $(2) - (1) - (2)$ & 18 \\
                  & & & $(1) - (2) - (1)$ & 12 \\
                  & & & $(3) - (0) - (1)$ &  8 \\
                  & & & $(1) - (0) - (3)$ &  8 \\
                  & & & $(2) - (1) - (0)$ &  6 \\
                  & & & $(0) - (1) - (2)$ &  6 \\
                  & & & $(1) - (0) - (1)$ &  4 \\
                  & & & $(0) - (1) - (0)$ &  2 \\[1ex] \hline
 \end{tabular}
 \end{center}
\end{table}

\section{The Search for Surviving Chains}

In the preceding section, we have described in some detail the arguments that
are needed to analyze the first step of the symmetry breaking process through
chains of subalgebras, during which the original supersymmetry is removed.
All further steps involve only ordinary Lie algebras and are carried out
according to the strategy already used in \cite{HH} and explained in detail
in \cite{HHF}; see also \cite{ABFH}. Briefly, the main criterion for excluding
a given chain without having to analyze all of its ramifications is the
occurence of one of the following situations:
\vspace{-2mm}
\begin{itemize}
 \item Total pairing: all multiplets come in pairs of equal or complex
       conjugate representations. No further breaking is able to remove
       this feature, excluding the possibility to produce multiplets with
       odd multiplicity, that is, the $3$ sextets, $5$ quartets and $9$
       doublets found in the genetic code.
 \item More than $2$ singlets. No further breaking is able to reduce
       the number of singlets, excluding the possibility to produce
       no more than the $2$ singlets found in the genetic code.
 \item More than $4$ odd-dimensional multiplets. No further breaking
       is able to reduce the number of odd-dimensional multiplets,
       excluding the possibility to produce no more than the $2$
       triplets and $2$ singlets found in the genetic code.
\vspace{-2mm}
\end{itemize}
In what follows, we list the chains that can be excluded by one of these
arguments, together with the relevant information on the distribution of
multiplets obtained after the last step.
\vspace{-2mm}
\begin{itemize}
 \item $A(2\,|\,0) = \mbox{\fraktw sl}(3\,|\,1)$: \\[1mm]
       Total pairing.
 \item $A(3\,|\,0) = \mbox{\fraktw sl}(4\,|\,1)$: \\[1mm]
       Continuing the symmetry breaking process, we obtain the following
       chains, all of which can be excluded:
 \begin{itemize}
  \item $A(3\,|\,0) \supset A_3 \supset A_2$:
        $10$ triplets and $6$ singlets.
  \item $A(3\,|\,0) \supset A_3 \supset C_2 \supset A_1 \oplus A_1$:
        $4$ triplets and $4$ singlets.
        
\pagebreak

  \item $A(3\,|\,0) \supset A_3 \supset C_2 \supset A_1$: \\
        $2$ septets, $2$ quintets, $2$ triplets and $2$ singlets.
  \item $A(3\,|\,0) \supset A_3 \supset A_1 \oplus A_1$:
        $2$ nonets, $4$ triplets and $2$ singlets.
 \end{itemize}
 \item $A(5\,|\,0) = \mbox{\fraktw sl}(6\,|\,1)$: \\[1mm]
       Continuing the symmetry breaking process, we obtain the following
       chains, all of which can be excluded:
 \begin{itemize}
  \item $A(5\,|\,0) \supset A_5 \supset A_4$:
        $4$ quintets and $4$ singlets, as well as total pairing.
  \item $A(5\,|\,0) \supset A_5 \supset A_3$:
        Total pairing.
  \item $A(5\,|\,0) \supset A_5 \supset C_3$:
        $4$ singlets.
  \item $A(5\,|\,0) \supset A_5 \supset A_2$:
        Total pairing.
  \item $A(5\,|\,0) \supset A_5 \supset A_1 \oplus A_3$:
        $4$ singlets.
  \item $A(5\,|\,0) \supset A_5 \supset A_2 \oplus A_2$:
        $4$ nonets, $8$ triplets and $4$ singlets.
  \item $A(5\,|\,0) \supset A_5 \supset A_1 \oplus A_2
                    \supset A_1 \oplus A_1^{(1)}$, where
        $\, A_2 \supset A_1^{(1)} \,$ corresponds to
        $\, \mbox{\fraktw su}(3) \supset \mbox{\fraktw su}(2)$:
        $4$ triplets and $4$ singlets.
  \item $A(5\,|\,0) \supset A_5 \supset A_1 \oplus A_2
                    \supset A_1 \oplus A_1^{(2)}$, where
        $\, A_2 \supset A_1^{(2)} \,$ corresponds to
        $\, \mbox{\fraktw su}(3) \supset \mbox{\fraktw so}(3)$:
        $2$ nonets, $2$ quintets and $4$ singlets.
 \end{itemize}
 \item $A(1\,|\,1)^c = \mbox{\fraktw sl}(2\,|\,2)$,
       highest weight $(1,l_2,1)$: \footnotemark
       \footnotetext{The superscript $.^c$ stands for ``central extension''.}
       \\[1mm]
       Too many odd-dimensional multiplets.
 \item $A(2\,|\,1) = \mbox{\fraktw sl}(3\,|\,2)$: \\[1mm]
       Continuing the symmetry breaking process, we obtain the following
       chains, all of which can be excluded:
 \begin{itemize}
  \item $A(2\,|\,1) \supset A_2 \oplus A_1 \supset A_1 \oplus A_1^{(1)}$,
        where $\, A_2 \supset A_1^{(1)} \,$ corresponds to
        $\, \mbox{\fraktw su}(3) \supset \mbox{\fraktw su}(2)$:
        $4$ triplets and $4$ singlets.
  \item $A(2\,|\,1) \supset A_2 \oplus A_1 \supset A_1 \oplus A_1^{(2)}$,
        where $\, A_2 \supset A_1^{(2)} \,$ corresponds to
        $\, \mbox{\fraktw su}(3) \supset \mbox{\fraktw so}(3)$:
        $2$ nonets, $2$ quintets and $4$ singlets.
 \end{itemize}
 \item $C(3) = \mbox{\fraktw osp}(2\,|\,4)$: \\[1mm]
       Continuing the symmetry breaking process, we obtain the following
       chains, all of which can be excluded:
 \begin{itemize}
  \item $C(3) \supset C_2 \supset A_1 \oplus A_1$:
        $4$ triplets and $4$ singlets.
  \item $C(3) \supset C_2 \supset A_1$: \\
        $2$ septets, $2$ quintets, $2$ triplets and $2$ singlets. \\
 \end{itemize}
 \item $C(4) = \mbox{\fraktw osp}(2\,|\,6)$: \\[1mm]
       Too many singlets.
\vspace{-2mm}
\end{itemize}
In the terminology of ref.\ \cite{HHF}, we are thus left with six basic
classical Lie super\-algebras whose codon representations, up to the end
of the first phase of the symmetry breaking process, produce surviving
chains: their remaining symmetry is described by a direct sum of
$\mbox{\fraktw sl}(2)$-algebras.

Finally, we must pass to the second phase of the symmetry breaking process,
during which some of the $\mbox{\fraktw sl}(2)$-algebras are broken. There
are two ways of doing this, depending on whether one uses the operator $L_z$
or the operator $L_z^2$ as the symmetry breaking term in the model Hamiltonian;
we shall in what follows refer to these two possibilities as ``strong''
breaking and ``soft'' breaking, respectively. However, only the first of
them corresponds to a genuine symmetry breaking at the level of Lie algebras,
namely from the Lie algebra $\mbox{\fraktw sl}(2)$ to its Cartan subalgebra.
A natural interpretation of both possibilities as a legitimate symmetry
breaking requires passing from the complex Lie algebra $\mbox{\fraktw sl}(2)$
to its compact real form $\mbox{\fraktw su}(2)$ and from there to the
corresponding connected, simply connected Lie group $SU(2)$, which all
have the same representation theory: then as has been observed in ref.\
\cite{FHH}, we may break the symmetry under the (connected) group $SU(2)$
in two different ways:~ a) down to its maximal connected subgroup $\, U(1)
\cong SO(2) \,$ (strong breaking) or~  b) down to its maximal (non-%
connected) subgroup $\, \mbox{\Bbbtw Z}_2 \times U(1) \cong O(2) \,$
(soft breaking). The effect on a multiplet of dimension $\, 2s+1$,
corresponding to an irreducible representation of $SU(2)$ (or
$\mbox{\fraktw su}(2)$ or $\mbox{\fraktw sl}(2)$) of spin $s$
and highest weight $2s$, is to break it~ a) strongly into
$\, 2s+1 \,$ singlets, corresponding to the different eigenvalues
of the operator $L_z$, or~ b) softly into
\vspace{-2mm}
\begin{itemize}
 \item $s$ doublets and one singlet if $s$ is integer, or
 \vspace{-1mm}
 \item $s$ doublets if $s$ is half-integer,
\vspace{-2mm}
\end{itemize}
corresponding to the different eigenvalues of the operator $L_z^2$.

The main complication in this second phase of the symmetry breaking process
arises from the necessity to take into account the possibility of (partially)
``freezing'' the symmetry breakdown in the last step; for more details, see
the discussion in ref.\ \cite{HHF}.

As an immediate consequence of the previous discussion, we see that the chain
resulting from the codon representation of $\mbox{\fraktw sl}(2\,|\,1)$ can
be excluded: all multiplets are of dimension $> 6$ so that further symmetry
breaking is needed (i.e., no freezing is allowed), but the remaining
symmetry algebra being a single copy of $\mbox{\fraktw sl}(2)$, any
further breaking will produce only singlets or doublets.

\pagebreak

The most stringent criterion for a chain to be surviving during the second
phase of the symmetry breaking process comes from the requirement of producing
the correct number of sextets (3) and triplets (2): it demands, among other
things, that the number
$$
 d_3~=~\parbox{7cm}{sum of the dimensions of all multiplets \\ \hspace*{\fill}
                    whose dimension is a multiple of 3 \hspace*{\fill}}
$$
which during this phase cannot decrease, must always remain $\, \geq 24$.
As an example, note that this condition immediately eliminates the codon
representation of $\mbox{\fraktw osp}(3\,|\,2)$, for which $\, d_3 = 18$,
according to \mbox{Table 3}. The remaining cases must be handled case by
case, as follows.
\addtocounter{footnote}{-1}
\begin{itemize}
 \item $A(1\,|\,1)^c = \mbox{\fraktw sl}(2\,|\,2)$,
       highest weight $(3,l_2,0)$. \footnotemark \\[1mm]
       Up to the end of the first phase, we have a unique chain:
       $$\mbox{\fraktw sl}(2\,|\,2) \supset
         \mbox{\fraktw sl}(2) \oplus \mbox{\fraktw sl}(2).
       $$
       The corresponding distribution of multiplets can be read off
       from \mbox{Table 2}; there are altogether $10$ multiplets, with
       $\, d_3 = 30$. However, among the four multiplets whose dimension
       is a multiple of $3$, we have one multiplet of dimension $6$, namely
       $(5)-(0)$, which cannot break into triplets, one multiplet of
       dimension $12$, namely $(3)-(2)$, which can either break into
       four triplets or else will produce no triplets at all, and
       finally two identical multiplets of dimension $6$, namely
       $(2)-(1)$, which together can also either break into four
       triplets or else produce no triplets at all. Thus there is
       no possibility to generate the two triplets found in the
       genetic code, so this chain may be discarded.
 \item $B(1\,|\,2) = \mbox{\fraktw osp}(3\,|\,4)$, highest weight
       $(0,{5 \over 2},3)$. \\[1mm]
       Up to the end of the first phase, we have the following chains.
 \begin{enumerate}
  \item $\mbox{\fraktw osp}(3\,|\,4) \supset
         \mbox{\fraktw sp}(4) \oplus \mbox{\fraktw so}(3) \supset
         \mbox{\fraktw sl}(2) \oplus \mbox{\fraktw sl}(2) \oplus
         \mbox{\fraktw sl}(2)$. \\[1mm]
        The corresponding distribution of multiplets can be read off
        from \mbox{Table 4}; there are altogether $8$ multiplets, with
        $\, d_3 = 24$. However, the two multiplets whose dimension is a
        multiple of $3$, namely $(1)-(0)-(5)$ and $(0)-(1)-(5)$, both of
        dimension $12$, cannot break into triplets, so this chain may
        be discarded.
  \item $\mbox{\fraktw osp}(3\,|\,4) \supset
         \mbox{\fraktw sp}(4) \oplus \mbox{\fraktw so}(3) \supset
         \mbox{\fraktw sl}(2) \oplus \mbox{\fraktw sl}(2)$. \\[1mm]
        The corresponding distribution of multiplets is identical with
        that shown in \mbox{Table 3}, since no further branching occurs
        in the second reduction; there are altogether $5$ multiplets,
        with $\, d_3 = 24$. However, the unique multiplet whose dimension
        is a multiple of $3$, namely $(3)-(5)$, of dimension $24$, cannot
        break into triplets, so this chain may be discarded.
 \end{enumerate}
       Continuing the first chain by diagonal breaking from three copies of
       $\mbox{\fraktw sl}(2)$ to two gives rise to the following additional
       chain.
 \begin{enumerate}
 \addtocounter{enumi}{2}
  \item $\mbox{\fraktw osp}(3\,|\,4) \supset
         \mbox{\fraktw sp}(4) \oplus \mbox{\fraktw so}(3) \supset
         \mbox{\fraktw sl}(2) \oplus \mbox{\fraktw sl}(2) \oplus
         \mbox{\fraktw sl}(2) \supset
         \mbox{\fraktw sl}(2)_{12} \oplus \mbox{\fraktw sl}(2)$. \\[1mm]
        The corresponding distribution of multiplets can be read off
        from \mbox{Table 4}; there are altogether $9$ multiplets, with
        $\, d_3 = 36$. However, among the three multiplets whose dimension
        is a multiple of $3$, we have two identical multiplets of dimension
        $12$, namely $(1)-(5)$, which cannot break into triplets, and one
        other multiplet of dimension $12$, namely $(2)-(3)$, which can
        either break into four triplets or else will produce no triplets
        at all. Thus there is no possibility to generate the two triplets
        found in the genetic code, so this chain may be discarded.
 \end{enumerate}
       The other possibilities of diagonal breaking by contracting the
       first or second $\mbox{\fraktw sl}(2)$ with the third can be ruled
       out because they lead to a total of $11$ multiplets where the number
       $d_3$ has already dropped to $21$, so there is no chance of producing
       the correct number of sextets and triplets.
\begin{table}[ht] \label{tb:BRTOSP34}
 \begin{center}
  Table 4: Branching of the codon representation of
           $\mbox{\fraktw osp}(3\,|\,4)$ (first phase)
 \end{center}
 \begin{center}
 \begin{tabular}{|c|c|c|c|c|c|} \hline
  \multicolumn{2}{|c|}{\rule{0mm}{5mm}
                       $\mbox{\fraktw sp}(4) \oplus \mbox{\fraktw so}(3)$
                             \rule{0mm}{5mm}} &
  \multicolumn{2}{|c|}{\rule{0mm}{5mm}
                       $\mbox{\fraktw sl}(2) \oplus \mbox{\fraktw sl}(2)
                                             \oplus \mbox{\fraktw sl}(2)$
                       \rule{0mm}{5mm}} &
  \multicolumn{2}{|c|}{\rule{0mm}{5mm}
                       $\mbox{\fraktw sl}(2)_{12} \oplus
                        \mbox{\fraktw sl}(2)$
                        \rule{0mm}{5mm}} \\[1mm] \hline
  {\footnotesize Highest Weight} &
  \rule{0mm}{5mm} {\footnotesize  $d$} \rule{0mm}{5mm} &
  {\footnotesize Highest Weight} &
  \rule{0mm}{5mm} {\footnotesize  $d$} \rule{0mm}{5mm} &
  {\footnotesize Highest Weight} &
  \rule{0mm}{5mm} {\footnotesize  $d$} \rule{0mm}{5mm} \\[1mm] \hline
  $(1,0)-(5)$ & $24$ & $(1)-(0)-(5)$ & $12$
        & \rule{0mm}{4.5mm} $(1)-(5)$ \rule{0mm}{4.5mm} & $12$ \\ \cline{3-6}
            &      & $(0)-(1)-(5)$ & $12$
        & \rule{0mm}{4.5mm} $(1)-(5)$ \rule{0mm}{4.5mm} & $12$ \\ \hline
  $(0,1)-(3)$ & $20$ & $(1)-(1)-(3)$ & $16$
        & \rule{0mm}{4.5mm} $(2)-(3)$ \rule{0mm}{4.5mm} & $12$ \\ \cline{5-6}
  & & & & \rule{0mm}{4.5mm} $(0)-(3)$ \rule{0mm}{4.5mm} &  $4$ \\ \cline{3-6}
            &      & $(0)-(0)-(3)$ &  $4$
        & \rule{0mm}{4.5mm} $(0)-(3)$ \rule{0mm}{4.5mm} &  $4$ \\ \hline
  $(1,0)-(1)$ &  $8$ & $(1)-(0)-(1)$ &  $4$
        & \rule{0mm}{4.5mm} $(1)-(1)$ \rule{0mm}{4.5mm} &  $4$ \\ \cline{3-6}
            &      & $(0)-(1)-(1)$ &  $4$
        & \rule{0mm}{4.5mm} $(1)-(1)$ \rule{0mm}{4.5mm} &  $4$ \\ \hline
  $(0,0)-(7)$ &  $8$ & $(0)-(0)-(7)$ &  $8$
        & \rule{0mm}{4.5mm} $(0)-(7)$ \rule{0mm}{4.5mm} &  $8$ \\ \hline
  $(0,0)-(3)$ &  $4$ & $(0)-(0)-(3)$ &  $4$
        & \rule{0mm}{4.5mm} $(0)-(3)$ \rule{0mm}{4.5mm} &  $4$ \\ \hline
  \multicolumn{2}{|c|}{\rule{0mm}{5mm} {\footnotesize $5$ subspaces}
                       \rule{0mm}{5mm}} &
  \multicolumn{2}{|c|}{\rule{0mm}{5mm} {\footnotesize $8$ subspaces}
                       \rule{0mm}{5mm}} &
  \multicolumn{2}{|c|}{\rule{0mm}{5mm} {\footnotesize $9$ subspaces}
                       \rule{0mm}{5mm}} \\[1mm] \hline
 \end{tabular}
 \end{center}
\end{table}
 \item $B(2\,|\,1) = \mbox{\fraktw osp}(5\,|\,2)$, highest weight
       $({5 \over 2},0,1)$. \\[1mm]
       Up to the end of the first phase, we have the following chains.
 \begin{enumerate}
  \item $\mbox{\fraktw osp}(5\,|\,2) \supset
         \mbox{\fraktw sp}(2) \oplus \mbox{\fraktw so}(5) \supset
         \mbox{\fraktw sl}(2) \oplus \mbox{\fraktw sl}(2) \oplus
         \mbox{\fraktw sl}(2)$. \\[1mm]
        The corresponding distribution of multiplets can be read off
        from \mbox{Table 5}; there are altogether $10$ multiplets, with
        $\, d_3 = 48$. Note also the symmetry of the distribution of
        multiplets under exchange of the second with the third
        $\mbox{\fraktw sl}(2)$. \\[1mm]
        In the first step, we must consider the following four options:
  \begin{itemize}
   \item[1.] breaking the first $\mbox{\fraktw sl}(2)$ softly generates
             $12$ multiplets \\ with $\, d_3 = 36$,
   \item[2.] breaking the first $\mbox{\fraktw sl}(2)$ strongly generates
             $18$ multiplets \\ with $\, d_3 = 36$,
   \item[3.] breaking the second $\mbox{\fraktw sl}(2)$ softly generates
             $13$ multiplets \\ with $\, d_3 = 30$,
   \item[4.] breaking the second $\mbox{\fraktw sl}(2)$ strongly generates
             precisely \\ $21$ multiplets with $\, d_3 = 30$.
  \end{itemize}
        Note that the last option leads to an an interesting scheme that comes
        close to the genetic code but is slightly different, with $3$ sextets,
        $5$ quartets, $4$ triplets, $5$ doublets and $4$ singlets. In the other
        three cases, the symmetry breaking process must proceed to the next
        stage, leading to the following options:

\pagebreak

  \begin{itemize}
   \item[1.1] breaking the first $\mbox{\fraktw sl}(2)$ down strongly
              generates $18$ multiplets \linebreak with $\, d_3 = 36$,
              so the symmetry breaking must continue and there can be
              no freezing at this stage, leading to the same situation
              as option 2 above,
   \item[1.2] breaking the second $\mbox{\fraktw sl}(2)$ softly generates
              $15$ multiplets \\ with $\, d_3 = 18$,
   \item[1.3] breaking the second $\mbox{\fraktw sl}(2)$ strongly generates
              $25$ multiplets \\ with $\, d_3 = 18$, \rule[-1mm]{0mm}{2mm}
   \item[2.1] breaking the second $\mbox{\fraktw sl}(2)$ softly generates
              $22$ multiplets \\ with $\, d_3 = 18$,
   \item[2.2] breaking the second $\mbox{\fraktw sl}(2)$ strongly generates
              $35$ multiplets \\ with $\, d_3 = 18$, \rule[-1mm]{0mm}{2mm}
   \item[3.1] breaking the first $\mbox{\fraktw sl}(2)$ softly generates
              $15$ multiplets \\ with $\, d_3 = 18$,
   \item[3.2] breaking the first $\mbox{\fraktw sl}(2)$ strongly generates
              $22$ multiplets \\ with $\, d_3 = 18$,
   \item[3.3] breaking the second $\mbox{\fraktw sl}(2)$ down strongly
              generates precisely \linebreak $21$ multiplets with
              $\, d_3 = 30$, leading to the same situation as \linebreak
              option 4 above,
   \item[3.4] breaking the third $\mbox{\fraktw sl}(2)$ softly generates
              $16$ multiplets \\ with $\, d_3 = 12$,
   \item[3.5] breaking the third $\mbox{\fraktw sl}(2)$ strongly generates
              $26$ multiplets \\ with $\, d_3 = 12$.
  \end{itemize}
        As before, options 1.2, 3.1 and 3.4 are excluded, whereas in the cases
        of options 1.3, 2.1, 2.2, 3.2 and 3.5, the symmetry breaking process
        must terminate, and we must take into account the possibility of
        freezing. However, the multiplets of dimension $> 6$ must not be
        frozen. As it turns out, it is impossible to generate the correct
        number of sextets ($3$), triplets ($2$) and singlets ($2$). In the
        cases of options 1.3 and 3.5, we must break the multiplet of dimension
        $12$ coming from the $(1-1-2)$ and can therefore generate at most $3$
        sextets or $2$ sextets and $2$ triplets. In the cases of options 2.1
        and 3.2 (which without freezing would produce the same distribution
        of multiplets), there is no possibility of generating triplets.
        Finally, in the case of option 2.2, breaking or freezing any
        combination of the two doublets coming from the $(1-1-0)$, the
        two doublets coming from the $(0-3-0)$ and the three doublets
        coming from the $(2-1-0)$ will generate $14$, $12$, $10$, $8$,
        $6$, $4$ or no singlets, but not $2$ singlets.

\pagebreak

  \item $\mbox{\fraktw osp}(5\,|\,2) \supset
         \mbox{\fraktw sp}(2) \oplus \mbox{\fraktw so}(5) \supset
         \mbox{\fraktw sl}(2) \oplus \mbox{\fraktw sl}(2)$. \\[1mm]
        The corresponding distribution of multiplets is easily obtained;
        there are altogether $7$ multiplets, with $\, d_3 = 30$. However,
        among the three multiplets whose dimension is a multiple of $3$,
        we have one multiplet of dimension $12$, namely $(1)-(5)$, and
        one multiplet of dimension $6$, namely $(0)-(5)$, both of which
        cannot break into triplets, and one other multiplet of dimension
        $12$, namely $(2)-(3)$, which can either break into four triplets
        or else will produce no triplets at all. Thus there is no possibility
        to generate the two triplets found in the genetic code, so this chain
        may be discarded.
 \end{enumerate}
       Continuing the first chain by diagonal breaking from three copies of
       $\mbox{\fraktw sl}(2)$ to two gives rise to the following additional
       chain.
 \begin{enumerate}
 \addtocounter{enumi}{2}
  \item $\mbox{\fraktw osp}(5\,|\,2) \supset
         \mbox{\fraktw sp}(2) \oplus \mbox{\fraktw so}(5) \supset
         \mbox{\fraktw sl}(2) \oplus \mbox{\fraktw sl}(2) \oplus
         \mbox{\fraktw sl}(2) \supset
         \mbox{\fraktw sl}(2)_{12} \oplus \mbox{\fraktw sl}(2)$. \\[1mm]
        The corresponding distribution of multiplets can be read off
        from \mbox{Table 5}; there are altogether $14$ multiplets, with
        $\, d_3 = 33$. \\[1mm]
        In the first step, we must consider the following four options:
  \begin{itemize}
   \item[1.] breaking the first $\mbox{\fraktw sl}(2)$ softly generates
             precisely $21$ multiplets \\ with $\, d_3 = 18$,
   \item[2.] breaking the first $\mbox{\fraktw sl}(2)$ strongly generates
             $35$ multiplets \\ with $\, d_3 = 18$,
   \item[3.] breaking the second $\mbox{\fraktw sl}(2)$ softly generates
             $18$ multiplets \\ with $\, d_3 = 24$,
   \item[4.] breaking the second $\mbox{\fraktw sl}(2)$ strongly generates
             $28$ multiplets \\ with $\, d_3 = 24$.
  \end{itemize}
        Note that the first option leads to an an interesting scheme that comes
        close to the genetic code but is slightly different, with $2$ sextets,
        $7$ quartets, $2$ triplets, $8$ doublets and $2$ singlets. In the cases
        of options 2 and 4, the symmetry breaking process must terminate, and
        we must take into account the possibility of freezing. However, the
        multiplet of dimension $9$ must not be frozen, so we get at least
        $3$ triplets and at least $6$ odd-dimensional multiplets. Therefore,
        the only possibility of continuing the symmetry breaking process
        is case 3, leading to the following options:
  \begin{itemize}
   \item[3.1] breaking the first $\mbox{\fraktw sl}(2)$ softly generates
              $26$ multiplets \\ with $\, d_3 = 0$,
   \item[3.2] breaking the first $\mbox{\fraktw sl}(2)$ strongly generates
              $42$ multiplets \\ with $\, d_3 = 0$, \rule[-1mm]{0mm}{2mm}
   \item[3.3] breaking the second $\mbox{\fraktw sl}(2)$ down strongly
              generates $28$ multiplets with $\, d_3 = 24$.
  \end{itemize}
        In all three cases, the symmetry breaking process must terminate,
        and we must take into account the possibility of freezing. However,
        the multiplet of dimension $8$ must not be frozen and will break
        either into $2$ quartets or into $4$ doublets. In all three cases,
        we are able to reproduce the genetic code, provided the freezing
        is chosen appropriately, as shown in Tables 6-8.
 \end{enumerate}
       The remaining possibility of diagonal breaking by contracting the
       second $\mbox{\fraktw sl}(2)$ with the third can be ruled out because
       it leads to a total of $14$ multiplets where the number $d_3$ has
       already dropped to $12$, so there is no chance of producing the
       correct number of sextets and triplets.
\begin{table}[ht] \label{tb:BRTOSP52A}
 \begin{center}
  Table 5: Branching of the codon representation of
           $\mbox{\fraktw osp}(5\,|\,2)$ (first phase)
 \end{center}
 \begin{center}
 \begin{tabular}{|c|c|c|c|c|c|} \hline
  \multicolumn{2}{|c|}{\rule{0mm}{5mm}
                       $\mbox{\fraktw sp}(2) \oplus \mbox{\fraktw so}(5)$
                       \rule{0mm}{5mm}} &
  \multicolumn{2}{|c|}{\rule{0mm}{5mm}
                       $\mbox{\fraktw sl}(2) \oplus \mbox{\fraktw sl}(2)
                                             \oplus \mbox{\fraktw sl}(2)$
                       \rule{0mm}{5mm}} &
  \multicolumn{2}{|c|}{\rule{0mm}{5mm}
                       $\mbox{\fraktw sl}(2)_{12} \oplus
                        \mbox{\fraktw sl}(2)$
                        \rule{0mm}{5mm}} \\[1mm] \hline
  {\footnotesize Highest Weight} &
  \rule{0mm}{5mm} {\footnotesize  $d$} \rule{0mm}{5mm} &
  {\footnotesize Highest Weight} &
  \rule{0mm}{5mm} {\footnotesize  $d$} \rule{0mm}{5mm} &
  {\footnotesize Highest Weight} &
  \rule{0mm}{5mm} {\footnotesize  $d$} \rule{0mm}{5mm} \\[1mm] \hline
  $(1)-(1,1)$ & $32$ & $(1)-(2)-(1)$ & $12$
        & \rule{0mm}{4.5mm} $(3)-(1)$ \rule{0mm}{4.5mm} & $8$ \\ \cline{5-6}
  & & & & \rule{0mm}{4.5mm} $(1)-(1)$ \rule{0mm}{4.5mm} & $4$ \\ \cline{3-6}
            &      & $(1)-(1)-(2)$ & $12$
        & \rule{0mm}{4.5mm} $(2)-(2)$ \rule{0mm}{4.5mm} & $9$ \\ \cline{5-6}
  & & & & \rule{0mm}{4.5mm} $(0)-(2)$ \rule{0mm}{4.5mm} & $3$ \\ \cline{3-6}
            &      & $(1)-(1)-(0)$ & $4$
        & \rule{0mm}{4.5mm} $(2)-(0)$ \rule{0mm}{4.5mm} & $3$ \\ \cline{5-6}
  & & & & \rule{0mm}{4.5mm} $(0)-(0)$ \rule{0mm}{4.5mm} & $1$ \\ \cline{3-6}
            &      & $(1)-(0)-(1)$ & $4$
        & \rule{0mm}{4.5mm} $(1)-(1)$ \rule{0mm}{4.5mm} & $4$ \\ \hline
  $(0)-(0,3)$ & $20$ & $(0)-(2)-(1)$ & $6$
        & \rule{0mm}{4.5mm} $(2)-(1)$ \rule{0mm}{4.5mm} & $6$ \\ \cline{3-6}
            &      & $(0)-(1)-(2)$ & $6$
        & \rule{0mm}{4.5mm} $(1)-(2)$ \rule{0mm}{4.5mm} & $6$ \\ \cline{3-6}
            &      & $(0)-(3)-(0)$ & $4$
        & \rule{0mm}{4.5mm} $(3)-(0)$ \rule{0mm}{4.5mm} & $4$ \\ \cline{3-6}
            &      & $(0)-(0)-(3)$ & $4$
        & \rule{0mm}{4.5mm} $(0)-(3)$ \rule{0mm}{4.5mm} & $4$ \\ \hline
  $(2)-(0,1)$ & $12$ & $(2)-(1)-(0)$ & $6$
        & \rule{0mm}{4.5mm} $(3)-(0)$ \rule{0mm}{4.5mm} & $4$ \\ \cline{5-6}
  & & & & \rule{0mm}{4.5mm} $(1)-(0)$ \rule{0mm}{4.5mm} & $2$ \\ \cline{3-6}
            &      & $(2)-(0)-(1)$ & $6$
        & \rule{0mm}{4.5mm} $(2)-(1)$ \rule{0mm}{4.5mm} & $6$ \\ \hline
  \multicolumn{2}{|c|}{\rule{0mm}{5mm} {\footnotesize  $3$ subspaces}
                       \rule{0mm}{5mm}} &
  \multicolumn{2}{|c|}{\rule{0mm}{5mm} {\footnotesize $10$ subspaces}
                       \rule{0mm}{5mm}} &
  \multicolumn{2}{|c|}{\rule{0mm}{5mm} {\footnotesize $14$ subspaces}
                       \rule{0mm}{5mm}} \\[1mm] \hline
 \end{tabular}
 \end{center}
\end{table}
\begin{table}[ht] \label{tb:BRTOSP52B}
 \begin{center}
  Table 6: Branching of the codon representation of
           $\mbox{\fraktw osp}(5\,|\,2)$ (second phase): \\
           First option
 \end{center}
 \begin{center}
 \begin{tabular}{|c|c|c|c|c|c|@{}c@{\hspace{6pt}}c|c|} \hline
  \multicolumn{2}{|c|}{\rule{0mm}{5mm}
                       $\mbox{\fraktw sl}(2) \oplus \mbox{\fraktw sl}(2)
                                             \oplus \mbox{\fraktw sl}(2)$
                       \rule{0mm}{5mm}} &
   \multicolumn{2}{c|}{\rule{0mm}{5mm}
                       $\mbox{\fraktw sl}(2)_{12} \oplus
                        \mbox{\fraktw sl}(2)$
                       \rule{0mm}{5mm}} &
   \multicolumn{2}{c|}{\rule{0mm}{5mm}
                       $L_{3,z}^2$
                       \rule{0mm}{5mm}} & &
   \multicolumn{2}{c|}{\rule{0mm}{5mm} \hspace{-10pt}
                       $(L_{12,z}^2,L_{3,z}^2)$
                       \rule{0mm}{5mm}} \\[1mm] \hline
  {\footnotesize $2s_1 - 2s_2 - 2s_3$} &
  \rule{0mm}{5mm} {\footnotesize $d$} \rule{0mm}{5mm} &
  {\footnotesize $2s_{12} - 2s_3$} &
  \rule{0mm}{5mm} {\footnotesize $d$} \rule{0mm}{5mm} &
  {\footnotesize $2s_{12} - 2m_3$} &
  \rule{0mm}{5mm} {\footnotesize $d$} \rule{0mm}{5mm} & &
  {\footnotesize $2m_{12} - 2m_3$} &
  \rule{0mm}{5mm} {\footnotesize $d$} \rule{0mm}{5mm} \\[1mm] \hline
  $1-2-1$ & $12$ & \rule{0mm}{4.5mm} $3-1$ \rule{0mm}{4.5mm} & $8$ &
  $3 - (\pm 1)$  & $8$ &               & $(\pm 3) - (\pm 1)$ & $4$ \\
  \cline{7-9}
          &      & \rule{0mm}{4.5mm}       \rule{0mm}{4.5mm} &     &
                 &     &               & $(\pm 1) - (\pm 1)$ & $4$ \\
  \cline{3-9}
          &      & \rule{0mm}{4.5mm} $1-1$ \rule{0mm}{4.5mm} & $4$ &
  $1 - (\pm 1)$  & $4$ &               & $(\pm 1) - (\pm 1)$ & $4$ \\
  \hline
  $1-1-2$ & $12$ & \rule{0mm}{4.5mm} $2-2$ \rule{0mm}{4.5mm} & $9$ &
  $2 - (\pm 2)$  & $6$ & \vline \vline & $(\pm 2) - (\pm 2)$ & $4$ \\
  \cline{7-9}
          &      & \rule{0mm}{4.5mm}       \rule{0mm}{4.5mm} &     &
                 &     & \vline \vline & $   0    - (\pm 2)$ & $2$ \\
  \cline{5-9}
          &      & \rule{0mm}{4.5mm}       \rule{0mm}{4.5mm} &     &
  $2 -    0   $  & $3$ & \vline \vline & $(\pm 2) -    0   $ & $2$ \\
  \cline{7-9}
          &      & \rule{0mm}{4.5mm}       \rule{0mm}{4.5mm} &     &
                 &     & \vline \vline & $   0    -    0   $ & $1$ \\
  \cline{3-9}
          &      & \rule{0mm}{4.5mm} $0-2$ \rule{0mm}{4.5mm} & $3$ &
  $0 - (\pm 2)$  & $2$ &               & $   0    - (\pm 2)$ & $2$ \\
  \cline{5-9}
          &      & \rule{0mm}{4.5mm}       \rule{0mm}{4.5mm} &     &
  $0 -    0   $  & $1$ &               & $   0    -    0   $ & $1$ \\
  \hline
  $1-1-0$ &  $4$ & \rule{0mm}{4.5mm} $2-0$ \rule{0mm}{4.5mm} & $3$ &
  $2 -    0   $  & $3$ & \vline \vline & $(\pm 2) -    0   $ & $2$ \\
  \cline{7-9}
          &      & \rule{0mm}{4.5mm}       \rule{0mm}{4.5mm} &     &
                 &     & \vline \vline & $   0    -    0   $ & $1$ \\
  \cline{3-9}
          &      & \rule{0mm}{4.5mm} $0-0$ \rule{0mm}{4.5mm} & $1$ &
  $0 -    0   $  & $1$ &               & $   0    -    0   $ & $1$ \\
  \hline
  $1-0-1$ & $4$  & \rule{0mm}{4.5mm} $1-1$ \rule{0mm}{4.5mm} & $4$ &
  $1 - (\pm 1)$  & $4$ &               & $(\pm 1) - (\pm 1)$ & $4$ \\
  \hline
  $0-2-1$ &  $6$ & \rule{0mm}{4.5mm} $2-1$ \rule{0mm}{4.5mm} & $6$ &
  $2 - (\pm 1)$  & $6$ & \vline \vline & $(\pm 2) - (\pm 1)$ & $4$ \\
  \cline{7-9}
          &      & \rule{0mm}{4.5mm}       \rule{0mm}{4.5mm} &     &
                 &     & \vline \vline & $   0    - (\pm 1)$ & $2$ \\
  \hline
  $0-1-2$ &  $6$ & \rule{0mm}{4.5mm} $1-2$ \rule{0mm}{4.5mm} & $6$ &
  $1 - (\pm 2)$  & $4$ &               & $(\pm 1) - (\pm 2)$ & $4$ \\
  \cline{5-9}
          &      & \rule{0mm}{4.5mm}       \rule{0mm}{4.5mm} &     &
  $1 -    0   $  & $2$ &               & $(\pm 1) -    0   $ & $2$ \\
  \hline
  $0-3-0$ &  $4$ & \rule{0mm}{4.5mm} $3-0$ \rule{0mm}{4.5mm} & $4$ &
  $3 -    0   $  & $4$ &               & $(\pm 3) -    0   $ & $2$ \\
  \cline{7-9}
          &      & \rule{0mm}{4.5mm}       \rule{0mm}{4.5mm} &     &
                 &     &               & $(\pm 1) -    0   $ & $2$ \\
  \hline
  $0-0-3$ &  $4$ & \rule{0mm}{4.5mm} $0-3$ \rule{0mm}{4.5mm} & $4$ &
  $0 - (\pm 3)$  & $2$ &               & $   0    - (\pm 3)$ & $2$ \\
  \cline{5-9}
          &      & \rule{0mm}{4.5mm}       \rule{0mm}{4.5mm} &     &
  $0 - (\pm 1)$  & $2$ &               & $   0    - (\pm 1)$ & $2$ \\
  \hline
  $2-1-0$ &  $6$ & \rule{0mm}{4.5mm} $3-0$ \rule{0mm}{4.5mm} & $4$ &
  $3 -    0   $  & $4$ &               & $(\pm 3) -    0   $ & $2$ \\
  \cline{7-9}
          &      & \rule{0mm}{4.5mm}       \rule{0mm}{4.5mm} &     &
                 &     &               & $(\pm 1) -    0   $ & $2$ \\
  \cline{3-9}
          &      & \rule{0mm}{4.5mm} $1-0$ \rule{0mm}{4.5mm} & $2$ &
  $1 -    0   $  & $2$ &               & $(\pm 1) -    0   $ & $2$ \\
  \hline
  $2-0-1$ &  $6$ & \rule{0mm}{4.5mm} $2-1$ \rule{0mm}{4.5mm} & $6$ &
  $2 - (\pm 1)$  & $6$ & \vline \vline & $(\pm 2) - (\pm 1)$ & $4$ \\
  \cline{7-9}
          &      & \rule{0mm}{4.5mm}       \rule{0mm}{4.5mm} &     &
                 &     & \vline \vline & $   0    - (\pm 1)$ & $2$ \\
  \hline
  \multicolumn{2}{|c|}{\rule{0mm}{5mm} {\footnotesize $10$ subspaces}
                       \rule{0mm}{5mm}} &
   \multicolumn{2}{c|}{\rule{0mm}{5mm} {\footnotesize $14$ subspaces}
                       \rule{0mm}{5mm}} &
   \multicolumn{2}{c|}{\rule{0mm}{5mm} {\footnotesize $18$ subspaces}
                       \rule{0mm}{5mm}} & &
   \multicolumn{2}{c|}{\rule{0mm}{5mm} \hspace{-10pt}
                                       {\footnotesize $26$ subspaces}
                       \rule{0mm}{5mm}} \\[1mm] \hline
 \end{tabular}
 \end{center}
\end{table}
\begin{table}[ht] \label{tb:BRTOSP52C}
 \begin{center}
  Table 7: Branching of the codon representation of
           $\mbox{\fraktw osp}(5\,|\,2)$ (second phase): \\
           Second option
 \end{center}
 \begin{center}
 \begin{tabular}{|c|c|c|c|c|c|@{}c@{\hspace{6pt}}c|c|} \hline
  \multicolumn{2}{|c|}{\rule{0mm}{5mm}
                       $\mbox{\fraktw sl}(2) \oplus \mbox{\fraktw sl}(2)
                                             \oplus \mbox{\fraktw sl}(2)$
                       \rule{0mm}{5mm}} &
   \multicolumn{2}{c|}{\rule{0mm}{5mm}
                       $\mbox{\fraktw sl}(2)_{12} \oplus
                        \mbox{\fraktw sl}(2)$
                       \rule{0mm}{5mm}} &
   \multicolumn{2}{c|}{\rule{0mm}{5mm}
                       $L_{3,z}^2$
                       \rule{0mm}{5mm}} & &
   \multicolumn{2}{c|}{\rule{0mm}{5mm} \hspace{-10pt}
                       $(L_{12,z},L_{3,z}^2)$
                       \rule{0mm}{5mm}} \\[1mm] \hline
  {\footnotesize $2s_1 - 2s_2 - 2s_3$} &
  \rule{0mm}{5mm} {\footnotesize $d$} \rule{0mm}{5mm} &
  {\footnotesize $2s_{12} - 2s_3$} &
  \rule{0mm}{5mm} {\footnotesize $d$} \rule{0mm}{5mm} &
  {\footnotesize $2s_{12} - 2m_3$} &
  \rule{0mm}{5mm} {\footnotesize $d$} \rule{0mm}{5mm} & &
  {\footnotesize $2m_{12} - 2m_3$} &
  \rule{0mm}{5mm} {\footnotesize $d$} \rule{0mm}{5mm} \\[1mm] \hline
  $1-2-1$ & $12$ & \rule{0mm}{4.5mm} $3-1$ \rule{0mm}{4.5mm} & $8$ &
  $3 - (\pm 1)$  & $8$ &               & $(+3) - (\pm 1)$ & $2$ \\
  \cline{7-9}
          &      & \rule{0mm}{4.5mm}       \rule{0mm}{4.5mm} &     &
                 &     &               & $(-3) - (\pm 1)$ & $2$ \\
  \cline{7-9}
          &      & \rule{0mm}{4.5mm}       \rule{0mm}{4.5mm} &     &
                 &     &               & $(+1) - (\pm 1)$ & $2$ \\
  \cline{7-9}
          &      & \rule{0mm}{4.5mm}       \rule{0mm}{4.5mm} &     &
                 &     &               & $(-1) - (\pm 1)$ & $2$ \\
  \cline{3-9}
          &      & \rule{0mm}{4.5mm} $1-1$ \rule{0mm}{4.5mm} & $4$ &
  $1 - (\pm 1)$  & $4$ & \vline \vline & $(+1) - (\pm 1)$ & $2$ \\
  \cline{7-9}
          &      & \rule{0mm}{4.5mm}       \rule{0mm}{4.5mm} &     &
                 &     & \vline \vline & $(-1) - (\pm 1)$ & $2$ \\
  \hline
  $1-1-2$ & $12$ & \rule{0mm}{4.5mm} $2-2$ \rule{0mm}{4.5mm} & $9$ &
  $2 - (\pm 2)$  & $6$ & \vline \vline & $(+2) - (\pm 2)$ & $2$ \\
  \cline{7-9}
          &      & \rule{0mm}{4.5mm}       \rule{0mm}{4.5mm} &     &
                 &     & \vline \vline & $(-2) - (\pm 2)$ & $2$ \\
  \cline{7-9}
          &      & \rule{0mm}{4.5mm}       \rule{0mm}{4.5mm} &     &
                 &     & \vline \vline & $ 0   - (\pm 2)$ & $2$ \\
  \cline{5-9}
          &      & \rule{0mm}{4.5mm}       \rule{0mm}{4.5mm} &     &
  $2 -    0   $  & $3$ & \vline \vline & $(+2) -    0   $ & $1$ \\
  \cline{7-9}
          &      & \rule{0mm}{4.5mm}       \rule{0mm}{4.5mm} &     &
                 &     & \vline \vline & $(-2) -    0   $ & $1$ \\
  \cline{7-9}
          &      & \rule{0mm}{4.5mm}       \rule{0mm}{4.5mm} &     &
                 &     & \vline \vline & $ 0   -    0   $ & $1$ \\
  \cline{3-9}
          &      & \rule{0mm}{4.5mm} $0-2$ \rule{0mm}{4.5mm} & $3$ &
  $0 - (\pm 2)$  & $2$ &               & $ 0   - (\pm 2)$ & $2$ \\
  \cline{5-9}
          &      & \rule{0mm}{4.5mm}       \rule{0mm}{4.5mm} &     &
  $0 -    0   $  & $1$ &               & $ 0   -    0   $ & $1$ \\
  \hline
  $1-1-0$ &  $4$ & \rule{0mm}{4.5mm} $2-0$ \rule{0mm}{4.5mm} & $3$ &
  $2 -    0   $  & $3$ & \vline \vline & $(+2) -    0   $ & $1$ \\
  \cline{7-9}
          &      & \rule{0mm}{4.5mm}       \rule{0mm}{4.5mm} &     &
                 &     & \vline \vline & $(-2) -    0   $ & $1$ \\
  \cline{7-9}
          &      & \rule{0mm}{4.5mm}       \rule{0mm}{4.5mm} &     &
                 &     & \vline \vline & $ 0   -    0   $ & $1$ \\
  \cline{3-9}
          &      & \rule{0mm}{4.5mm} $0-0$ \rule{0mm}{4.5mm} & $1$ &
  $0 -    0   $  & $1$ &               & $ 0   -    0   $ & $1$ \\
  \hline
  $1-0-1$ & $4$  & \rule{0mm}{4.5mm} $1-1$ \rule{0mm}{4.5mm} & $4$ &
  $1 - (\pm 1)$  & $4$ & \vline \vline & $(+1) - (\pm 1)$ & $2$ \\
  \cline{7-9}
          &      & \rule{0mm}{4.5mm}       \rule{0mm}{4.5mm} &     &
                 &     & \vline \vline & $(-1) - (\pm 1)$ & $2$ \\
  \hline
  \multicolumn{9}{|c|}{\rule{0mm}{5mm}
                       {\footnotesize Table 7 continued on next page}
                       \rule{0mm}{5mm}} \\[1mm]
 \end{tabular}
 \end{center}
\end{table}
\begin{table}[ht] \label{tb:BRTOSP52D}
 \begin{center}
 \begin{tabular}{|c|c|c|c|c|c|@{}c@{\hspace{6pt}}c|c|}
  \multicolumn{9}{|c|}{\rule{0mm}{5mm}
                       {\footnotesize Table 7 continued from previous page}
                       \rule{0mm}{5mm}} \\[1mm] \hline
  $0-2-1$ &  $6$ & \rule{0mm}{4.5mm} $2-1$ \rule{0mm}{4.5mm} & $6$ &
  $2 - (\pm 1)$  & $6$ & \vline \vline & $(+2) - (\pm 1)$ & $2$ \\
  \cline{7-9}
          &      & \rule{0mm}{4.5mm}       \rule{0mm}{4.5mm} &     &
                 &     & \vline \vline & $(-2) - (\pm 1)$ & $2$ \\
  \cline{7-9}
          &      & \rule{0mm}{4.5mm}       \rule{0mm}{4.5mm} &     &
                 &     & \vline \vline & $ 0   - (\pm 1)$ & $2$ \\
  \hline
  $0-1-2$ &  $6$ & \rule{0mm}{4.5mm} $1-2$ \rule{0mm}{4.5mm} & $6$ &
  $1 - (\pm 2)$  & $4$ & \vline \vline & $(+1) - (\pm 2)$ & $2$ \\
  \cline{7-9}
          &      & \rule{0mm}{4.5mm}       \rule{0mm}{4.5mm} &     &
                 &     & \vline \vline & $(-1) - (\pm 2)$ & $2$ \\
  \cline{5-9}
          &      & \rule{0mm}{4.5mm}       \rule{0mm}{4.5mm} &     &
  $1 -    0   $  & $2$ & \vline \vline & $(+1) -    0   $ & $1$ \\
  \cline{7-9}
          &      & \rule{0mm}{4.5mm}       \rule{0mm}{4.5mm} &     &
                 &     & \vline \vline & $(-1) -    0   $ & $1$ \\
  \hline
  $0-3-0$ &  $4$ & \rule{0mm}{4.5mm} $3-0$ \rule{0mm}{4.5mm} & $4$ &
  $3 -    0   $  & $4$ & \vline \vline & $(+3) -    0   $ & $1$ \\
  \cline{7-9}
          &      & \rule{0mm}{4.5mm}       \rule{0mm}{4.5mm} &     &
                 &     & \vline \vline & $(-3) -    0   $ & $1$ \\
  \cline{7-9}
          &      & \rule{0mm}{4.5mm}       \rule{0mm}{4.5mm} &     &
                 &     & \vline \vline & $(+1) -    0   $ & $1$ \\
  \cline{7-9}
          &      & \rule{0mm}{4.5mm}       \rule{0mm}{4.5mm} &     &
                 &     & \vline \vline & $(-1) -    0   $ & $1$ \\
  \hline
  $0-0-3$ &  $4$ & \rule{0mm}{4.5mm} $0-3$ \rule{0mm}{4.5mm} & $4$ &
  $0 - (\pm 3)$  & $2$ &               & $ 0   - (\pm 3)$ & $2$ \\
  \cline{5-9}
          &      & \rule{0mm}{4.5mm}       \rule{0mm}{4.5mm} &     &
  $0 - (\pm 1)$  & $2$ &               & $ 0   - (\pm 1)$ & $2$ \\
  \hline
  $2-1-0$ &  $6$ & \rule{0mm}{4.5mm} $3-0$ \rule{0mm}{4.5mm} & $4$ &
  $3 -    0   $  & $4$ & \vline \vline & $(+3) -    0   $ & $1$ \\
  \cline{7-9}
          &      & \rule{0mm}{4.5mm}       \rule{0mm}{4.5mm} &     &
                 &     & \vline \vline & $(-3) -    0   $ & $1$ \\
  \cline{7-9}
          &      & \rule{0mm}{4.5mm}       \rule{0mm}{4.5mm} &     &
                 &     & \vline \vline & $(+1) -    0   $ & $1$ \\
  \cline{7-9}
          &      & \rule{0mm}{4.5mm}       \rule{0mm}{4.5mm} &     &
                 &     & \vline \vline & $(-1) -    0   $ & $1$ \\
  \cline{3-9}
          &      & \rule{0mm}{4.5mm} $1-0$ \rule{0mm}{4.5mm} & $2$ &
  $1 -    0   $  & $2$ & \vline \vline & $(+1) -    0   $ & $1$ \\
  \cline{7-9}
          &      & \rule{0mm}{4.5mm}       \rule{0mm}{4.5mm} &     &
                 &     & \vline \vline & $(-1) -    0   $ & $1$ \\
  \hline
  $2-0-1$ &  $6$ & \rule{0mm}{4.5mm} $2-1$ \rule{0mm}{4.5mm} & $6$ &
  $2 - (\pm 1)$  & $6$ & \vline \vline & $(+2) - (\pm 1)$ & $2$ \\
  \cline{7-9}
          &      & \rule{0mm}{4.5mm}       \rule{0mm}{4.5mm} &     &
                 &     & \vline \vline & $(-2) - (\pm 1)$ & $2$ \\
  \cline{7-9}
          &      & \rule{0mm}{4.5mm}       \rule{0mm}{4.5mm} &     &
                 &     & \vline \vline & $ 0   - (\pm 1)$ & $2$ \\
  \hline
  \multicolumn{2}{|c|}{\rule{0mm}{5mm} {\footnotesize $10$ subspaces}
                       \rule{0mm}{5mm}} &
   \multicolumn{2}{c|}{\rule{0mm}{5mm} {\footnotesize $14$ subspaces}
                       \rule{0mm}{5mm}} &
   \multicolumn{2}{c|}{\rule{0mm}{5mm} {\footnotesize $18$ subspaces}
                       \rule{0mm}{5mm}} & &
   \multicolumn{2}{c|}{\rule{0mm}{5mm} \hspace{-10pt}
                                       {\footnotesize $42$ subspaces}
                       \rule{0mm}{5mm}} \\[1mm] \hline
 \end{tabular}
 \end{center}
\end{table}
\begin{table}[ht] \label{tb:BRTOSP52E}
 \begin{center}
  Table 8: Branching of the codon representation of
           $\mbox{\fraktw osp}(5\,|\,2)$ (second phase): \\
           Third option
 \end{center}
 \begin{center}
 \begin{tabular}{|c|c|c|c|c|c|@{}c@{\hspace{6pt}}c|c|} \hline
  \multicolumn{2}{|c|}{\rule{0mm}{5mm}
                       $\mbox{\fraktw sl}(2) \oplus \mbox{\fraktw sl}(2)
                                             \oplus \mbox{\fraktw sl}(2)$
                       \rule{0mm}{5mm}} &
   \multicolumn{2}{c|}{\rule{0mm}{5mm}
                       $\mbox{\fraktw sl}(2)_{12} \oplus
                        \mbox{\fraktw sl}(2)$
                       \rule{0mm}{5mm}} &
   \multicolumn{2}{c|}{\rule{0mm}{5mm}
                       $L_{3,z}^2$
                       \rule{0mm}{5mm}} & &
   \multicolumn{2}{c|}{\rule{0mm}{5mm} \hspace{-10pt}
                       $L_{3,z}$
                       \rule{0mm}{5mm}} \\[1mm] \hline
  {\footnotesize $2s_1 - 2s_2 - 2s_3$} &
  \rule{0mm}{5mm} {\footnotesize $d$} \rule{0mm}{5mm} &
  {\footnotesize $2s_{12} - 2s_3$} &
  \rule{0mm}{5mm} {\footnotesize $d$} \rule{0mm}{5mm} &
  {\footnotesize $2s_{12} - 2m_3$} &
  \rule{0mm}{5mm} {\footnotesize $d$} \rule{0mm}{5mm} & &
  {\footnotesize $2s_{12} - 2m_3$} &
  \rule{0mm}{5mm} {\footnotesize $d$} \rule{0mm}{5mm} \\[1mm] \hline
  $1-2-1$ & $12$ & \rule{0mm}{4.5mm} $3-1$ \rule{0mm}{4.5mm} & $8$ &
  $3 - (\pm 1)$  & $8$ &               & $3 - (+1)$ & $4$ \\ \cline{7-9}
          &      & \rule{0mm}{4.5mm}       \rule{0mm}{4.5mm} &     &
                 &     &               & $3 - (-1)$ & $4$ \\ \cline{3-9}
          &      & \rule{0mm}{4.5mm} $1-1$ \rule{0mm}{4.5mm} & $4$ &
  $1 - (\pm 1)$  & $4$ &               & $1 - (+1)$ & $2$ \\ \cline{7-9}
          &      & \rule{0mm}{4.5mm}       \rule{0mm}{4.5mm} &     &
                 &     &               & $1 - (-1)$ & $2$ \\ \hline
  $1-1-2$ & $12$ & \rule{0mm}{4.5mm} $2-2$ \rule{0mm}{4.5mm} & $9$ &
  $2 - (\pm 2)$  & $6$ & \vline \vline & $2 - (+2)$ & $3$ \\ \cline{7-9}
          &      & \rule{0mm}{4.5mm}       \rule{0mm}{4.5mm} &     &
                 &     & \vline \vline & $2 - (-2)$ & $3$ \\ \cline{5-9}
          &      & \rule{0mm}{4.5mm}       \rule{0mm}{4.5mm} &     &
  $2 -    0   $  & $3$ & \vline \vline & $2 -  0  $ & $3$ \\ \cline{3-9}
          &      & \rule{0mm}{4.5mm} $0-2$ \rule{0mm}{4.5mm} & $3$ &
  $0 - (\pm 2)$  & $2$ & \vline \vline & $0 - (+2)$ & $1$ \\ \cline{7-9}
          &      & \rule{0mm}{4.5mm}       \rule{0mm}{4.5mm} &     &
                 &     & \vline \vline & $0 - (-2)$ & $1$ \\ \cline{5-9}
          &      & \rule{0mm}{4.5mm}       \rule{0mm}{4.5mm} &     &
  $0 -    0   $  & $1$ & \vline \vline & $0 -  0  $ & $1$ \\ \hline
  $1-1-0$ &  $4$ & \rule{0mm}{4.5mm} $2-0$ \rule{0mm}{4.5mm} & $3$ &
  $2 -    0   $  & $3$ &               & $2 -  0  $ & $3$ \\ \cline{3-9}
          &      & \rule{0mm}{4.5mm} $0-0$ \rule{0mm}{4.5mm} & $1$ &
  $0 -    0   $  & $1$ &               & $0 -  0  $ & $1$ \\ \hline
  $1-0-1$ & $4$  & \rule{0mm}{4.5mm} $1-1$ \rule{0mm}{4.5mm} & $4$ &
  $1 - (\pm 1)$  & $4$ &               & $1 - (+1)$ & $2$ \\ \cline{7-9}
          &      & \rule{0mm}{4.5mm}       \rule{0mm}{4.5mm} &     &
                 &     &               & $1 - (-1)$ & $2$ \\ \hline
  $0-2-1$ &  $6$ & \rule{0mm}{4.5mm} $2-1$ \rule{0mm}{4.5mm} & $6$ &
  $2 - (\pm 1)$  & $6$ & \vline \vline & $2 - (+1)$ & $3$ \\ \cline{7-9}
          &      & \rule{0mm}{4.5mm}       \rule{0mm}{4.5mm} &     &
                 &     & \vline \vline & $2 - (-1)$ & $3$ \\ \hline
  $0-1-2$ &  $6$ & \rule{0mm}{4.5mm} $1-2$ \rule{0mm}{4.5mm} & $6$ &
  $1 - (\pm 2)$  & $4$ & \vline \vline & $1 - (+2)$ & $2$ \\ \cline{7-9}
          &      & \rule{0mm}{4.5mm}       \rule{0mm}{4.5mm} &     &
                 &     & \vline \vline & $1 - (-2)$ & $2$ \\ \cline{5-9}
          &      & \rule{0mm}{4.5mm}       \rule{0mm}{4.5mm} &     &
  $1 -    0   $  & $2$ & \vline \vline & $1 -  0  $ & $2$ \\ \hline
  $0-3-0$ &  $4$ & \rule{0mm}{4.5mm} $3-0$ \rule{0mm}{4.5mm} & $4$ &
  $3 -    0   $  & $4$ &               & $3 -  0  $ & $4$ \\ \hline
  $0-0-3$ &  $4$ & \rule{0mm}{4.5mm} $0-3$ \rule{0mm}{4.5mm} & $4$ &
  $0 - (\pm 3)$  & $2$ & \vline \vline & $0 - (+3)$ & $1$ \\ \cline{7-9}
          &      & \rule{0mm}{4.5mm}       \rule{0mm}{4.5mm} &     &
                 &     & \vline \vline & $0 - (-3)$ & $1$ \\ \cline{5-9}
          &      & \rule{0mm}{4.5mm}       \rule{0mm}{4.5mm} &     &
  $0 - (\pm 1)$  & $2$ & \vline \vline & $0 - (+1)$ & $1$ \\ \cline{7-9}
          &      & \rule{0mm}{4.5mm}       \rule{0mm}{4.5mm} &     &
                 &     & \vline \vline & $0 - (-1)$ & $1$ \\ \hline
  $2-1-0$ &  $6$ & \rule{0mm}{4.5mm} $3-0$ \rule{0mm}{4.5mm} & $4$ &
  $3 -    0   $  & $4$ &               & $3 -  0  $ & $4$ \\ \cline{3-9}
          &      & \rule{0mm}{4.5mm} $1-0$ \rule{0mm}{4.5mm} & $2$ &
  $1 -    0   $  & $2$ &               & $1 -  0  $ & $2$ \\ \hline
  $2-0-1$ &  $6$ & \rule{0mm}{4.5mm} $2-1$ \rule{0mm}{4.5mm} & $6$ &
  $2 - (\pm 1)$  & $6$ & \vline \vline & $2 - (+1)$ & $3$ \\ \cline{7-9}
          &      & \rule{0mm}{4.5mm}       \rule{0mm}{4.5mm} &     &
                 &     & \vline \vline & $2 - (-1)$ & $3$ \\ \hline
  \multicolumn{2}{|c|}{\rule{0mm}{5mm} {\footnotesize $10$ subspaces}
                       \rule{0mm}{5mm}} &
   \multicolumn{2}{c|}{\rule{0mm}{5mm} {\footnotesize $14$ subspaces}
                       \rule{0mm}{5mm}} &
   \multicolumn{2}{c|}{\rule{0mm}{5mm} {\footnotesize $18$ subspaces}
                       \rule{0mm}{5mm}} & &
   \multicolumn{2}{c|}{\rule{0mm}{5mm} \hspace{-10pt}
                                       {\footnotesize $28$ subspaces}
                       \rule{0mm}{5mm}} \\[1mm] \hline
 \end{tabular}
 \end{center}
\end{table}
 \item $D(2\,|\,1) = \mbox{\fraktw osp}(4\,|\,2)$, highest weight
       $(5,0,0)$. \\[1mm]
       Up to the end of the first phase, we have a unique chain:
 \begin{enumerate}
  \item $\mbox{\fraktw osp}(4\,|\,2) \supset
         \mbox{\fraktw sl}(2) \oplus \mbox{\fraktw sl}(2) \oplus
         \mbox{\fraktw sl}(2)$. \\[1mm]
        The corresponding distribution of multiplets can be read off
        from \mbox{Table 3}; there are altogether $6$ multiplets, with
        $\, d_3 = 42$. Note also the symmetry of the distribution
        of multiplets under exchange of the second with the third
        $\mbox{\fraktw sl}(2)$. However, among the four multiplets
        whose dimension is a multiple of $3$, we have one multiplet
        of dimension $6$, namely $(5)-(0)-(0)$, which cannot break
        into triplets, and three multiplets of dimension $12$, namely
        $(3)-(2)-(0)$, $(3)-(0)-(2)$ and $(2)-(1)-(1)$, each of which
        can either break into four triplets or else will produce no
        triplets at all. Thus there is no possibility to generate
        the two triplets found in the genetic code, so this chain
        may be discarded.
 \end{enumerate}
       Continuing this chain by diagonal breaking from three copies of
       $\mbox{\fraktw sl}(2)$ to two gives rise to the following additional
       chains.
 \begin{enumerate}
 \addtocounter{enumi}{1}
  \item $\mbox{\fraktw osp}(4\,|\,2) \supset
         \mbox{\fraktw sl}(2) \oplus \mbox{\fraktw sl}(2) \oplus
         \mbox{\fraktw sl}(2) \supset
         \mbox{\fraktw sl}(2)_{12} \oplus \mbox{\fraktw sl}(2)$. \\[1mm]
        The corresponding distribution of multiplets is easily obtained;
        there are altogether $10$ multiplets, with $\, d_3 = 36$. However,
        among the four multiplets whose dimension is a multiple of $3$,
        we have one multiplet of dimension $12$, namely $(5)-(1)$, and
        two identical multiplets of dimension $6$, namely $(5)-(0)$,
        all of which cannot break into triplets, and one other multiplet
        of dimension $12$, namely $(3)-(2)$, which can either break into
        four triplets or else will produce no triplets at all. Thus
        there is no possibility to generate the two triplets found
        in the genetic code, so this chain may be discarded.

\pagebreak

  \item $\mbox{\fraktw osp}(4\,|\,2) \supset
         \mbox{\fraktw sl}(2) \oplus \mbox{\fraktw sl}(2) \oplus
         \mbox{\fraktw sl}(2) \supset
         \mbox{\fraktw sl}(2) \oplus \mbox{\fraktw sl}(2)_{23}$. \\[1mm]
        The corresponding distribution of multiplets can be read off
        from \mbox{Table 9}; there are altogether $8$ multiplets, with
        $\, d_3 = 57$. \\[1mm]
        In the first step, we must consider the following four options:
  \begin{itemize}
   \item[1.] breaking the first $\mbox{\fraktw sl}(2)$ softly generates
             $18$ multiplets \\ with $\, d_3 = 48$,
   \item[2.] breaking the first $\mbox{\fraktw sl}(2)$ strongly generates
             $32$ multiplets \\ with $\, d_3 = 48$,
   \item[3.] breaking the second $\mbox{\fraktw sl}(2)$ softly generates
             $12$ multiplets \\ with $\, d_3 = 18$,
   \item[4.] breaking the second $\mbox{\fraktw sl}(2)$ strongly generates
             $16$ multiplets \\ with $\, d_3 = 18$.
  \end{itemize}
        As before, options 3 and 4 are excluded, whereas in the case of
        option 2, the symmetry breaking process must terminate, and we
        must take into account the possibility of freezing. However,
        the multiplets of dimension $> 6$ and of dimension $5$ must not
        be frozen, so we get at least $16$ triplets and $5$ singlets.
        Therefore, the only possibility of continuing the symmetry breaking
        process is case 1, leading to the following options:
  \begin{itemize}
   \item[1.1] breaking the first $\mbox{\fraktw sl}(2)$ down strongly
              generates $32$ multiplets \\ with $\, d_3 = 48$.
   \item[1.2] breaking the second $\mbox{\fraktw sl}(2)$ softly generates
              $27$ multiplets \\ with $\, d_3 = 0$,
   \item[1.3] breaking the second $\mbox{\fraktw sl}(2)$ strongly generates
              $35$ multiplets \\ with $\, d_3 = 0$.
  \end{itemize}
        In all three cases, the symmetry breaking process must terminate,
        and we must take into account the possibility of freezing. However,
        we already have $2$ triplets and $2$ singlets at the previous stage,
        and the requirement that no new triplets or singlets may be generated
        forces the large majority of the multiplets to be frozen. As it turns
        out, it is possible to generate the correct number of sextets ($3$),
        triplets ($2$) and singlets ($2$), but not of quartets ($5$) and
        doublets ($9$); we get at most $4$ quartets and at least $11$
        doublets.
 \end{enumerate}
\begin{table}[ht] \label{tb:BRTOSP42}
 \begin{center}
  Table 9: Branching of the codon representation of
           $\mbox{\fraktw osp}(4\,|\,2)$ \\
           with highest weight $(5,0,0)$ (first phase)
 \end{center}
 \begin{center}
 \begin{tabular}{|c|c|c|c|} \hline
  \multicolumn{2}{|c|}{\rule{0mm}{5mm}
                       $\mbox{\fraktw sl}(2) \oplus \mbox{\fraktw sl}(2)
                                             \oplus \mbox{\fraktw sl}(2)$
                       \rule{0mm}{5mm}} &
  \multicolumn{2}{|c|}{\rule{0mm}{5mm}
                       $\mbox{\fraktw sl}(2) \oplus
                        \mbox{\fraktw sl}(2)_{23}$
                        \rule{0mm}{5mm}} \\[1mm] \hline
  {\footnotesize Highest Weight} &
  \rule{0mm}{5mm} {\footnotesize  $d$} \rule{0mm}{5mm} &
  {\footnotesize Highest Weight} &
  \rule{0mm}{5mm} {\footnotesize  $d$} \rule{0mm}{5mm} \\[1mm] \hline
  $(4)-(1)-(1)$ & $20$
    & \rule{0mm}{4.5mm} $(4)-(2)$ \rule{0mm}{4.5mm} & $15$ \\ \cline{3-4}
  & & \rule{0mm}{4.5mm} $(4)-(0)$ \rule{0mm}{4.5mm} &  $5$ \\ \hline
  $(3)-(2)-(0)$ & $12$
    & \rule{0mm}{4.5mm} $(3)-(2)$ \rule{0mm}{4.5mm} & $12$ \\ \hline
  $(3)-(0)-(2)$ & $12$
    & \rule{0mm}{4.5mm} $(3)-(2)$ \rule{0mm}{4.5mm} & $12$ \\ \hline
  $(2)-(1)-(1)$ & $12$
    & \rule{0mm}{4.5mm} $(2)-(2)$ \rule{0mm}{4.5mm} & $9$ \\ \cline{3-4}
  & & \rule{0mm}{4.5mm} $(2)-(0)$ \rule{0mm}{4.5mm} & $3$ \\ \hline
  $(5)-(0)-(0)$ &  $6$
    & \rule{0mm}{4.5mm} $(5)-(0)$ \rule{0mm}{4.5mm} & $6$ \\ \hline
  $(1)-(0)-(0)$ &  $2$
    & \rule{0mm}{4.5mm} $(1)-(0)$ \rule{0mm}{4.5mm} & $2$ \\ \hline
  \multicolumn{2}{|c|}{\rule{0mm}{5mm} {\footnotesize $6$ subspaces}
                       \rule{0mm}{5mm}} &
  \multicolumn{2}{|c|}{\rule{0mm}{5mm} {\footnotesize $8$ subspaces}
                       \rule{0mm}{5mm}} \\[1mm] \hline
 \end{tabular}
 \end{center}
\end{table}
 \item $D(2\,|\,1) = \mbox{\fraktw osp}(4\,|\,2)$, highest weight
       $({7 \over 2},0,1)$. \\[1mm]
       Up to the end of the first phase, we have a unique chain:
 \begin{enumerate}
  \item $\mbox{\fraktw osp}(4\,|\,2) \supset
         \mbox{\fraktw sl}(2) \oplus \mbox{\fraktw sl}(2) \oplus
         \mbox{\fraktw sl}(2)$. \\[1mm]
        The corresponding distribution of multiplets can be read off
        from \mbox{Table 3}; there are altogether $8$ multiplets, with
        $\, d_3 = 42$. Note also the symmetry of the distribution of
        multiplets under exchange of the first with the third
        $\mbox{\fraktw sl}(2)$. \\[1mm]
        In the first step, we must consider the following four options:
  \begin{itemize}
   \item[1.] breaking the first $\mbox{\fraktw sl}(2)$ softly generates
             $11$ multiplets \\ with $\, d_3 = 36$,
   \item[2.] breaking the first $\mbox{\fraktw sl}(2)$ strongly generates
             $18$ multiplets \\ with $\, d_3 = 36$,
   \item[3.] breaking the second $\mbox{\fraktw sl}(2)$ softly generates
             $9$ multiplets \\ with $\, d_3 = 30$,
   \item[4.] breaking the second $\mbox{\fraktw sl}(2)$ strongly generates
             $14$ multiplets \\ with $\, d_3 = 30$, but among them are
             $2$ nonets, $4$ triplets and \\ $2$ singlets.
  \end{itemize}
        In the first three cases, the symmetry breaking process must proceed
        to the next stage, leading to the following options:
  \begin{itemize}
   \item[1.1] breaking the first $\mbox{\fraktw sl}(2)$ down strongly
              generates $18$ multiplets \linebreak with $\, d_3 = 36$,
              so the symmetry breaking must continue and there can be
              no freezing at this stage, leading to the same situation
              as option 2 above,
   \item[1.2] breaking the second $\mbox{\fraktw sl}(2)$ softly generates
              $12$ multiplets \\ with $\, d_3 = 24$,
   \item[1.3] breaking the second $\mbox{\fraktw sl}(2)$ strongly generates
              $19$ multiplets \\ with $\, d_3 = 24$, but among them are
              $4$ triplets and $4$ singlets,
   \item[1.4] breaking the third $\mbox{\fraktw sl}(2)$ softly generates
              $15$ multiplets \\ with $\, d_3 = 12$,
   \item[1.5] breaking the third $\mbox{\fraktw sl}(2)$ strongly generates
              $24$ multiplets \\ with $\, d_3 = 12$, \rule[-1mm]{0mm}{2mm}
   \item[2.1] breaking the second $\mbox{\fraktw sl}(2)$ softly generates
              $20$ multiplets \\ with $\, d_3 = 24$,
   \item[2.2] breaking the second $\mbox{\fraktw sl}(2)$ strongly generates
              $30$ multiplets \\ with $\, d_3 = 24$,
   \item[2.3] breaking the third $\mbox{\fraktw sl}(2)$ softly generates
              $24$ multiplets \\ with $\, d_3 = 12$,
   \item[2.4] breaking the third $\mbox{\fraktw sl}(2)$ strongly generates
              $40$ multiplets \\ with $\, d_3 = 12$, \rule[-1mm]{0mm}{2mm}
   \item[3.1] breaking the first $\mbox{\fraktw sl}(2)$ softly generates
              $12$ multiplets \\ with $\, d_3 = 24$, leading to the same
              situation as option 1.2,
   \item[3.2] breaking the first $\mbox{\fraktw sl}(2)$ strongly generates
              $20$ multiplets \\ with $\, d_3 = 24$, leading to the same
              situation as option 2.1,
   \item[3.3] breaking the second $\mbox{\fraktw sl}(2)$ down strongly
              generates $14$ multiplets with $\, d_3 = 30$, leading to the
              same situation as option 4 above.
  \end{itemize}
  \pagebreak
        As before, option 1.4 is excluded, whereas in the case of options
        1.5, 2.2, 2.3 and 2.4, the symmetry breaking process must terminate,
        and we must take into account the possibility of freezing. However,
        the multiplets of dimension $> 6$ must not be frozen. In the cases
        of options 1.5 and 2.3, we do not get any triplets or singlets at
        all. In the case of option 2.4, we either do not get any triplets
        or singlets at all or else we get too many (at least $4$). In the
        case of option 2.2, we are able to produce the correct number of
        sextets ($3$), triplets ($2$) and singlets ($2$), but there is no
        possibility to generating the correct number of quartets ($5$) and
        doublets ($9$): we can only get $2$ quartets and $15$ doublets.
        In the case of option 2.1, we already have $20$ multiplets but
        no triplets and no singlets: their generation would require
        breaking at least two multiplets in the next step (one sextet
        and one doublet, for example), leading to at least $22$ multiplets.
        We are thus left with a single surviving option for continuing the
        symmetry breaking process, namely $1.2\!=\!3.1$, which consists in
        breaking both the first and the second $\mbox{\fraktw sl}(2)$ softly,
        generating $12$ multiplets with $\, d_3 = 24$, giving rise to the
        following options:
  \begin{itemize}
   \item[a)] breaking the first $\mbox{\fraktw sl}(2)$ down strongly
             generates $20$ multiplets \\ with $\, d_3 = 24$, leading to
             the same situation as option 2.1 above,
   \item[b)] breaking the second $\mbox{\fraktw sl}(2)$ down strongly
             generates $19$ multiplets \\ with $\, d_3 = 24$, leading to
             the same situation as option 1.3 above,
   \item[c)] breaking the third $\mbox{\fraktw sl}(2)$ softly generates
             $16$ multiplets \\ with $\, d_3 = 0$,
   \item[d)] breaking the third $\mbox{\fraktw sl}(2)$ strongly generates
             $26$ multiplets \\ with $\, d_3 = 0$.
  \end{itemize}
        As before, option c) is excluded, whereas in the case of option d),
        we do not get any triplets or singlets at all.
 \end{enumerate}
       Continuing this chain by diagonal breaking from three copies of
       $\mbox{\fraktw sl}(2)$ to two gives rise to the following additional
       chain.
 \begin{enumerate}
 \addtocounter{enumi}{1}
  \item $\mbox{\fraktw osp}(4\,|\,2) \supset
         \mbox{\fraktw sl}(2) \oplus \mbox{\fraktw sl}(2) \oplus
         \mbox{\fraktw sl}(2) \supset
         \mbox{\fraktw sl}(2)_{12} \oplus \mbox{\fraktw sl}(2)$. \\[1mm]
        The corresponding distribution of multiplets is easily obtained;
        there are altogether $11$ multiplets, with $\, d_3 = 24$. However,
        among the three multiplets whose dimension is a multiple of $3$,
        we have one multiplet of dimension $12$, namely $(3)-(2)$, which
        can either break into four triplets or else will produce no triplets
        at all, and two identical multiplets of dimension $6$, namely
        $(1)-(2)$, which together can also either break into four
        triplets or else produce no triplets at all. Thus there is
        no possibility to generate the two triplets found in the
        genetic code, so this chain may be discarded.
 \end{enumerate}
       The remaining possibility of diagonal breaking by contracting the
       first $\mbox{\fraktw sl}(2)$ with the third can be ruled out because
       it leads to a total of $14$ multiplets among which there are $1$
       nonet, $2$ quintets, $4$ triplets and $1$ singlet.
\end{itemize}

\section{Conclusion}

The main results of the analysis presented in ref.\ \cite{FS} and in the
present paper, which in preliminary form were announced in \cite{FS1} and
\cite{FS2}, can be summarized as follows.

The idea of describing the degeneracies of the genetic code as the result of
a symmetry breaking process through chains of subalgebras can be investigated
systematically within the context of typical codon representations of basic
classical Lie superalgebras, instead of ordinary codon representations of
ordinary simple Lie algebras. The first result is negative: as before,
there is no symmetry breaking pattern through chains of subalgebras
capable of reproducing exactly the degeneracies of the genetic code.
In other words, the phenomenon of ``freezing'' remains an essential part
of the approach. The second result is positive and, as far as the uniqueness
part is concerned, more stringent than its non-supersymmetric counterpart:
admitting the possibility of ``freezing'' during the last step of the
procedure, we find three schemes that do reproduce the degeneracies
of the standard code, all based on the ortho\-symplectic algebra
$\mbox{\fraktw osp}(5|2)$ and differing only in the detailed form
of the symmetry breaking pattern during the last step. \linebreak
The most natural scheme, shown in Tables 5 and 6, is the one that
allows for a simple choice of Hamiltonian, in the sense used in
ref.\ \cite{HH} and explained in more detail in ref.\ \cite{HHF},
namely the following:
\begin{equation}
 \begin{array}{rcl}
  H~=~H_0 \!\!\!&+&\!\!\! \lambda \, C_2(\mbox{\fraktw so}(5)) \, + \,
                          \alpha_1 \bm{L}_1^2 \, + \,
                          \alpha_2 \bm{L}_2^2 \, + \,
                          \alpha_3 \bm{L}_3^2 \, + \,
                          \alpha_{12} (\bm{L}_1 + \bm{L}_2)^2 \\[1mm]
                &+&\!\!\! \beta_3 L_{3,z}^2 \, + \,
                          \gamma_{12} \, ((\bm{L}_1 + \bm{L}_2)^2 - 2) \,
                          (L_{1,z} + L_{2,z})^2~.
 \end{array}
\end{equation} 
The investigation of the resulting $\mbox{\fraktw osp}(5\,|\,2)$ model for
the genetic code is presently under way.

\subsection*{Acknowledgments}

The authors would like to thank Prof.\ J.E.M.~Hornos for his incentive and
support of the present project, Prof.\ A.~Sciarrino and Prof.\ P.~Jarvis for
clarifying correspondence on the representation theory of Lie superalgebras
and Prof.\ \mbox{A.\ Grishkov} for fruitful discussions.

\pagebreak


\begin{thebibliography}{99}

\bibitem{HH} J.E.M.~Hornos and Y.M.M.~Hornos: {\em Algebraic Model for the
 Evolution of the Genetic Code}, Phys.\ Rev.\ Lett.\ {\bf 71} (1993) 4401-4404.

\bibitem{K1} V.G.~Kac: {\em Lie Superalgebras}, Adv.\ Math.\ {\bf 26}
 (1977) 8-96.

\bibitem{K2} V.G.~Kac: {\em Representations of Classical Lie Superalgebras},
 in: Proceedings of the VIth International Conference on Differential
 Geometric Methods in Theoretical Physics, Bonn, Germany 1977, Lecture
 Notes in Mathematics, Vol.\ 676, pp.\ 597-626, Springer, Berlin (1978). 

\bibitem{FS} M.~Forger and S.~Sachse: {\em Lie Superalgebras and the Multiplet
 Structure of the Genetic Code I: Codon Representations}, Preprint RT-MAP 9802,
 October 1998.

\bibitem{HHF} J.E.M.~Hornos, Y.M.M.~Hornos and M.~Forger: {\em Symmetry and
 Symmetry Breaking: An Algebraic Approach to the Genetic Code}, Preprint,
 April 1999.

\bibitem{Boe} H.~Boerner: {\em Representations of Groups}, North Holland,
 Amsterdam (1970).

\bibitem{Ham} M.~Hamermesh: {\em Group Theory and its Application to Physical
 Problems}, Addison-Wesley, Reading (1962).

\bibitem{MSS} B.~Morel, A.~Sciarrino and P.~Sorba: {\em Representations of
 $OSp\,(M|2n)$ and Young Supertableaux}, J.\ Phys.\ A: Math.\ Gen.\ {\bf 18}
 (1985) 1597-1613.

\bibitem{CK} C.J.~Cummins and R.C.~King: {\em Young Diagrams, Supercharacters
 of $OSp\,(M\,|\,N)$ and Modification Rules}, J.\ Phys.\ A: Math.\ Gen.\
 {\bf 20} (1987) 3103-3120, {\em Composite Young Diagrams, Supercharacters
 of $U(M\,|\,N)$ and Modification Rules}, J.\ Phys.\ A: Math.\ Gen.\
 {\bf 20} (1987) 3121-3133.

\bibitem{BMR} I.~Bars, B.~Morel and H.~Ruegg: {\em Kac-Dynkin Diagrams
 and Supertableaux}, J.\ Math.\ Phys.\ {\bf 24} (1983) 2253-2262.

\bibitem{BB} A.B.~Balantekin and I.~Bars: {\em Branching Rules for the
 Supergroup $SU(N\,|\,M)$ from those of $SU(N+M)$}, J.\ Math.\ Phys.\
 {\bf 22} (1981) 1239-1247.

\bibitem{IN} C.~Itzykson and M.~Nauenberg: {\em Unitary Groups:
 Representations and Decompositions}, Rev.\ Mod.\ Phys.\ {\bf 38}
 (1966) 95-120.

\bibitem{GSS1} G.~Girardi, A.~Sciarrino and P.~Sorba: {\em Kronecker Products
 for $SO(2p)$ Representations}, J.\ Phys.\ A: Math.\ Gen.\ {\bf 15} (1982)
 1119-1129.

\bibitem{GSS2} G.~Girardi, A.~Sciarrino and P.~Sorba: {\em Kronecker Product
 of $Sp\,(2n)$ Representations Using Generalized Young Tableaux}, J.\ Phys.\ A:
 Math.\ Gen.\ {\bf 16} (1983) 2609-2614.

\bibitem{VJ} J.~Van der Jeugt: {\em Irreducible Representations of the
 Exceptional Lie Superalgebras $D(2\,|\,1;\alpha)$}, J.\ Math.\ Phys.\
 {\bf 26} (1985) 913-924.

\bibitem{ABFH} F.~Antoneli~Jr., L.~Braggion, M.~Forger and J.E.M.~Hornos:
 {\em Extending the Search for Symmetries in the Genetic Code},
 in preparation.

\bibitem{FHH} M.~Forger, Y.M.M.~Hornos and J.E.M.~Hornos:
 {\em Global Aspects in the Algebraic Approach to the Genetic Code},
 Phys.\ Rev.\ {\bf E 56} (1997) 7078-7082.

\bibitem{FS1} S.~Sachse and M.~Forger: {\em Lie Superalgebras and the Multiplet
 Structure of the Genetic Code}, in: Proceedings of the 5th Wigner Symposium,
 Wien, Austria (1997), pp.\ 254-256, eds.: P.~Kasperkovitz and D.~Grau,
 World Scientific, Singapore (1998).

\bibitem{FS2} S.~Sachse and M.~Forger: {\em An Orthosymplectic Symmetry
 for the Genetic Code}, in: Group 22 -- Proceedings of the XXIInd
 International Colloquium on Group Theoretical Methods in Physics,
 Hobart, Australia 1998, pp.\ 147-151, eds.: S.P.~Corney, R.~Delbourgo
 and P.D.~Jarvis, International Press, Boston (1999).

\end{thebibliography}
\end{document}